\documentclass[11pt]{article}
\usepackage{amssymb}
\usepackage{graphics,graphicx}
\usepackage{color}
\usepackage{amsbsy}
\usepackage[left=2cm,right=2cm,top=2cm,bottom=2cm,bindingoffset=0cm]{geometry}
\usepackage{amsmath}
\usepackage{ccaption}
\usepackage{epstopdf}
\usepackage{verbatim}
\usepackage{dsfont}
\usepackage{soul}

\usepackage[affil-it]{authblk}

\begin{document}

\title{Analysis of dynamic failure of the discrete chain structure with non-local interactions.}

\author{Nikolai Gorbushin%
  \thanks{Corresponding author. Room 501, Department of Mathematics, IMPACS, Aberystwyth University, Ceredigion, SY23~3BZ, UK. Email: \texttt{nig15@aber.ac.uk}}}
\author{Gennady Mishuris}
\affil{Department of Mathematics, IMPACS, Aberystwyth University, Ceredigion, SY23~3BZ, UK}

\date{}
\maketitle

\begin{abstract}
In the present work the steady-state crack propagation in a chain of oscillators with non-local interactions is considered. The interactions are modeled as linear springs while the crack is presented by the absence of extra springs. The problem is reduced to the Wiener-Hopf type and solution is presented in terms of the inverse Fourier transform. It is shown that the non-local interactions may change the structure of the solution well-known from the classical local interactions formulation.
In particular, it may change the range of the region of stable crack motion. The conclusions of the analysis are supported by numerical results.
Namely, the observed phenomenon is partially clarified by evaluation of the structure profiles on the crack line ahead.
\end{abstract}

\maketitle

\vspace{-6pt}
\section{Introduction}

Classical linear elasticity theory describes various phenomena in homogeneous solids and composites under different loading conditions and has been effectively used in engineering applications. However, sometimes there is a necessity in utilizing more complex elastic continua models when describing stress concentration in essentially heterogeneous materials or when dynamic problems in the materials are considered for the wavelengths which are much larger than the characteristic size of heterogeneity. All those reasons gave rise for the appearance different theories which are able to overcome above mentioned restrictions of the classic elasticity. Among them one can mention the non-local elasticity ~\cite{eringen1972}, higher order strain gradient theory ~\cite{mindlin1964} or Cosserat continuum ~\cite{cosserat1968,mindlin1962}. The first one assumes non-local relationship between the stress and strain tensors, the second takes into account the higher derivatives of strain field in stress-strain equations. The Cosserat theory takes into consideration not only the translation of points as in classical elasticity but also their rotation, not only a stress (a force per unit area) but also a couple stress (a torque per unit area). The strong connection of these theories with the discrete lattice with non-local interactions can be found in numerous papers ~\cite{dipaola2010,polyzos2012,rosenau1987,suiker2001,tarasov2015}.  Extensive overview of the different lattice models are presented in \cite{ostojastarzewski2002}. The link between high order strain theories and discrete models is discussed in ~\cite{andrianov2010,askes2002}. The recent work \cite{michelitsch2014} shows the non-locality effect leading to general non-local constitutive relations of materials in the absence of defects. Wave propagation in a dispersive non-local model of a periodic composite and composites reinforced by continuous fibres have been considered by Federico Sabina and co-authors~\cite{vivar2009,levin2011}.

Discrete lattice model with non-local interactions were used in \cite{trofimov2010,truskinovsky2005} to analyse phase transitions. The authors demonstrated an impact of the non-local field on character of the propagation of the phase transition. The authors of~\cite{kresse2003} used bistable springs to describe the lattice defect movement in a discrete chain. They utilized a technique allowing them to avoid factorization problem. Moreover, since the solution was written directly in form of specific series, the construction of the deformation profiles was straightforward in contrast with the accurate technique involving Fourier transform and consequent solution of an auxiliary Wiener-Hopf problem as it was done for a similar problem in~\cite{slepyan2004,cherkaev2005}. Note that the phase transition problems although manifesting important dynamic features (propagation of the transition front)
do not involve the processes of the structural destruction.

Although there are numerous papers on wave propagation and phase transition in discrete lattice models with non-local interactions, surprisingly, they are not so many theoretical
studies on damage propagation in the structures. Recent numerical studies using N-th neighbour links in the discrete model ~\cite{chen2014} revealed that introduction of non-local interactions allows to reduce the crack path preference while the computational time increases with the number of interactions.

In the present paper we consider a crack propagation in a discrete structure with non-local interactions. In order to solve our problem we apply the Wiener-Hopf technique based on the ideas proposed by Slepyan~\cite{slepyan1984} and developed further in number of publications, e.g.~\cite{mishuris2012,mishuris2007,mishuris2008,mishuris2009,slepyan2005,slepyan2012,slepyan2004,slepyan2010}
for the local interactions taking place in various lattice structures. The crack propagation is assumed to be steady-state, i.e. the crack tip moves at a constant speed.
We show that the methods proved to be efficient for lattice structures with the local interaction are equally effective for these more complex structures. Additionally, we overcome the usual computational challenges related to accurate factorization and then inversion of the integral transform. Here we essentially used information on asymptotic behaviour of the functions in
singular points. This allows us to quantify the solution of the problem not only in terms of the energy diagrams but also through reconstructing  the displacement. That, in turn, gives us a chance to look closely on the existence and stability of the steady-state regimes.

\section{Problem formulation}

\begin{figure}[h!]
\center{\includegraphics[scale=0.7]{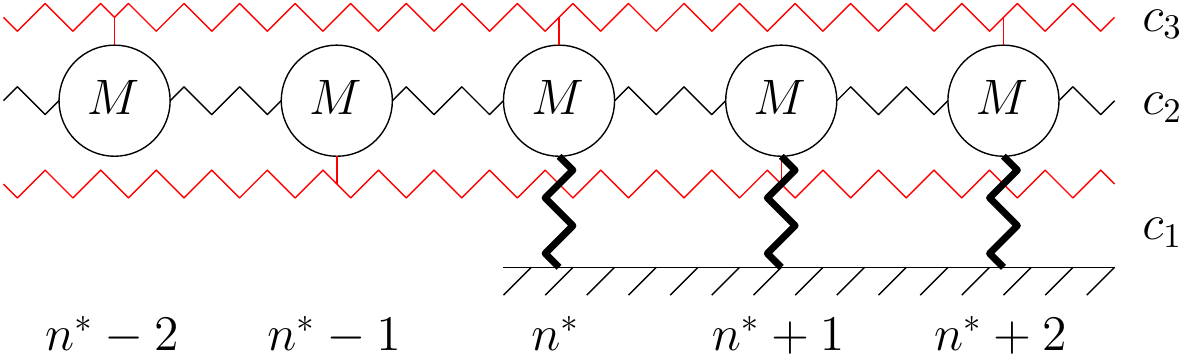}}
\caption[ ]{Chain of oscillators with equal masses $M$ connected to a substrate with springs $c_1$ (fat lines), closest neighbours with springs $c_2$ (normal lines), and second closest neighbours with springs $c_3$ (red lines), $n^*$ represents the crack tip. }
\label{Chain}
\end{figure}

Let us consider a discrete model with non-local interactions represented by a chain of masses and analyse the crack propagation in the structure presented in fig. \ref{Chain}.
The following notations are introduced here: $M$ is a mass of a oscillator, $c_1$ is a spring constant of the links between the oscillators and a substrate, $c_2$ is a stiffness of the bonds between the closest neighbours, $c_3$ is a spring constant between the second closest neighbouring masses.
We suppose that particles are evenly distributed with a separation of length $a$ and the vertical springs have also the same equilibrium length. Note that this configuration represents also the symmetric chain.

The coordinate $n^*(t)=Vt$ represents a location of the crack tip which propagates with a constant speed $V$ from left to right.
We study the effect of introduced non-local interactions, i.e. the magnitude of $c_3$ in the classical chain model described by the spring stiffness  $c_1$ and $c_2$.

Let us normalise the quantities by the equilibrium separation $a$:
\begin{equation}
u_n(t)=\frac{x_n(t)}{a},\quad v=\frac{V}{a},
\end{equation}
where $x_n(t)$ is the horizontal displacement of $n$-th oscillator from the equilibrium. Equations of motion for the mechanical system under consideration take form:
\begin{equation}
\begin{gathered}
M\frac{d^2u_n}{dt^2}=c_3(u_{n+2}+u_{n-2}-2u_n)+c_2(u_{n+1}+u_{n-1}-2u_n)-c_1u_n, \quad  n\geq{n^*},\\
M\frac{d^2u_n}{dt^2}=c_3(u_{n+2}+u_{n-2}-2u_n)+c_2(u_{n+1}+u_{n-1}-2u_n), \quad n<n^*,
\end{gathered}
\label{EquationOfMotion}
\end{equation}

In the present work we consistently use the following deformation fracture criterion:
\begin{equation}
\begin{cases}
u_{n^*}=u_c,\\
u_{n}<u_c,\quad n>n^*,
\end{cases}
\label{FractureCriterion}
\end{equation}
where $u_c$ is a critical value of the normalised displacement field at the crack tip. Note that a more natural choice for the criterion would be the critical magnitude of the absolute value of displacement of the oscillator $|u_{n^*}|$ at the crack tip. However, such condition is not always consistent with the assumption on the steady-state crack propagation. We refer prospective reader to the paper~\cite{mishuris2009} where both criteria are considered and discussed. The condition \eqref{FractureCriterion} allows one to find the loading conditions which are determined in the further analysis.

\section{Solution of the problem}

We additionally assume that the crack propagates from the left to the right with a constant subsonic speed $v$, e.g.
\begin{equation}
v<v_c,\quad v_c^2=\frac{c_2+4c_3}{M},
\end{equation}
where $v_c$ is a speed of sound in the broken region of the chain ($n<n^*$) normalised by equilibrium separation $a$. It is convenient for the further analysis to introduce to the corresponding moving coordinate system by the relation
\begin{equation}
\eta=n-vt.
\end{equation}

Following \cite{slepyan2012} we assume that the new variable $\eta$ can be considered as a continuous variable even though $n$ is an integer. We suppose that the spatial variable $\eta$ and time $t$ can be treated as independent. The introduction of $\eta$ makes the broken region of a chain to be defined by values $\eta<0$ whereas the region $\eta>0$ remains intact. Considering the problem in a steady-state crack propagation regime the equations of motion \eqref{EquationOfMotion} are written in the form:
\begin{equation}
\begin{gathered}
Mv^2\frac{d^2}{d\eta^2}u(\eta)=c_3(u(\eta+2)+u(\eta-2)-2u(\eta))+c_2(u(\eta+1)+u(\eta-1)-2u(\eta))-c_1u(\eta)H(\eta),
\end{gathered}
\label{EqOfMotEta}
\end{equation}
where $H(\eta)$ is a Heaviside function. To obtain the solution to this reformulated problem we apply Fourier transform which leads to the equation:
\begin{equation}
M(0+ikv)^2U(k)=-4c_3\sin^2{k}U(k)-4c_2\sin^2{\left(\frac{k}{2}\right)}U(k)-c_1U^+(k),
\label{FourierTrans}
\end{equation}
where:
\begin{equation}
\begin{gathered}
U(k)=\int\limits_{-\infty}^{\infty}u(\eta)e^{ik\eta}\,d\eta=U^+(k)+U^-(k), \\
U^+(k)=\int\limits_{0}^{\infty}u(\eta)e^{ik\eta}\,d\eta, \quad U^-(k)=\int\limits_{-\infty}^{0}u(\eta)e^{ik\eta}\,d\eta.
\end{gathered}
\label{FourierTrans_1}
\end{equation}

From now on the subscript "$\pm$" of the function shows that the function is analytical in the half plane $\pm\text{Im}k>0$ while the expression $(0+ikv)$ should be understood as:
\begin{equation}
(0+ikv)=\lim_{s\to+0}(s+ikv),
\label{LimitSteadyState}
\end{equation}

The problem \eqref{FourierTrans} can be written in the form:
\begin{equation}
U^-(k)+L(k)U^+(k)=0,
\label{Wiener-Hopf}
\end{equation}
with
\begin{equation}
L(k)=1+\frac{c_1}{M(0+ikv)^2+4c_3\sin^2{k}+4c_2\sin^2{\left(\dfrac{k}{2}\right)}}.
\label{FunL}
\end{equation}

The complex valued function $\text{Log}L(k)=\log{|L(k)|}+i\text{Arg}L(k)$ possesses the following properties:
\begin{equation}
\begin{gathered}
L(k)=\overline{L(-k)},\\
|L(k)|=|L(-k)|,\quad \text{Arg}L(k)=-\text{Arg}L(-k),\quad \text{for } k\in\mathds{R},
\end{gathered}
\label{PropertiesL}
\end{equation}
and has the asymptotics at zero and infinity as follows:
\begin{equation}
L(k)\sim\frac{c_1}{M(v_c^2-v^2)}\frac{1}{(0+ik)(0-ik)},\quad k\to0,
\label{estim_0}
\end{equation}
\begin{equation}
L(k)=1-\frac{c_1}{Mk^2v^2}+O(k^{-4}),\quad  k\to\infty,
\label{estim_infty}
\end{equation}

Note that for any $s>0$ in \eqref{LimitSteadyState} and the properties \eqref{PropertiesL}, one can prove that the index of the function $L(x)$ on the real axis is equal to zero. Moreover,there are no roots along the real axis.
In the limiting case, $s\to+0$, the roots and poles of the function can move towards real axis. As a result, when integrating along the axis in \eqref{FourierTrans_1},
the integration path should be conventionally deformed to avoid zeros and poles on the real axis from the appropriate side (see \cite{slepyan2012}).

Utilising the properties \eqref{PropertiesL}, one can factorise function $L(k)$ in the standard way \cite{gakhov2014,kisil2015}:
\begin{equation}
\begin{gathered}
L(k)=L^+(k)L^-(k),\quad
L^\pm(k)=\exp{\left(\pm\frac{1}{2\pi i}\int\limits_{-\infty}^{\infty}\frac{\text{Log}{L(\xi)}}{\xi-k}\,d\xi\right)},
\end{gathered}
\label{Factorization}
\end{equation}
where the integration path should be taken in accordance with the comment above. Functions $L^\pm(k)$ are analytical in the corresponding half-spaces $\pm\text{Im}{k}>0$ (above and below the integration path in the limiting case). Moreover one can directly check that the factors also satisfy the conditions \eqref{PropertiesL} and thus $L^\pm(-k)=\overline{L^\pm(k)}$ for any $k\in\mathds{R}$.

Factorisation formulae \eqref{Factorization} may lead to a significant computational challenge when the singular points of $\text{Log} |L(k)|$ are located  on the real axis while the the integration path is respectively deformed. To avoid the mentioned difficulties, following \cite{slepyan2012}, we use a different factorisation method presented in Appendices \ref{AppendixDispersion} and \ref{AppendixFactorisation} (specifically, see \eqref{FunLDisp},\eqref{ZerosPoles},\eqref{Function_l+-},\eqref{FactorizationFinal}). Note also that the factorisation of the function $L(k)$ under the conditions is unique and does not naturally depend on the used technique.

Using the properties of the factors in \eqref{Factorization} one can conclude that $\text{Arg}L(k)$ is an odd function while $\log{|L(k)|}$ is an even function for $k\in\mathds{R}$. This in turns allows us to obtain asymptotic expansion of the factors $L^{\pm}(k)$ at infinity:
\begin{equation}
\begin{gathered}
L^{\pm}(k)-1\sim \pm i\frac{Q}{k},\quad k\to\infty,\\
Q=\frac{1}{\pi}\int\limits_{0}^{\infty}\text{log}{|L(\xi)|}\,d\xi.
\end{gathered}
\label{Asym_Inf_L+-}
\end{equation}

Similarly to \cite{slepyan2012} one can also estimate asymptotic behaviour of the factors near zero point with the use of Sokhotski-Plemelj relation:
\begin{equation}
\begin{gathered}
L^{\pm}(k)\sim R^{\pm1}\sqrt{\frac{c_1}{M(v_c^2-v^2)}}\frac{1}{(0\mp ik)}\left(1+(0\mp ik)S\right),\quad k\to0,\\[3mm]
R=\exp{\left(\frac{1}{\pi}\int\limits_0^\infty\frac{\text{Arg}L(\xi)}{\xi}\,d\xi\right)},\quad S=\frac{1}{\pi}\int\limits_0^\infty\frac{\log{|L(\xi)|}}{\xi^2}\,d\xi.
\end{gathered}
\label{R}
\end{equation}

After the factorisation, the Wiener-Hopf type problem \eqref{Wiener-Hopf} can be rewritten in the form:
\begin{equation}
\frac{U^-(k)}{L^-(k)}+L^+(k)U^+(k)=0,
\label{WienerFact}
\end{equation}
which possesses only trivial solutions as the equation does not incorporate any external load applied to the structure.

We assume that the propagation of the crack at the constant speed $v$ is generated by the energy supply (feeding energy) coming from $\eta=-\infty$ and has to be implemented into the model. Following ~\cite{slepyan2012} this can be achieved by the modification of equation \eqref{WienerFact} as:
\begin{equation}
\frac{U^-(k)}{L^-(k)}+L^+(k)U^+(k)=\frac{C}{0- ik}+\frac{C}{0+ ik},
\label{WienerFactModified}
\end{equation}
with the solution given in the form:
\begin{equation}
U^+(k)=\frac{1}{L^+(k)}\frac{C}{0- ik},\quad
U^-(k)=L^-(k)\frac{C}{0+ ik},
\label{SolutionFourier}
\end{equation}
where an unknown constant $C$ is to be specified. The expressions \eqref{Asym_Inf_L+-},\eqref{R},\eqref{SolutionFourier} allow one to obtain the asymptotic expressions for the sought-for functions $U^{\pm}(k)$:
\begin{equation}
\begin{gathered}
U^{\pm}(k)\sim C\left(\pm\frac{i}{k}+\frac{Q}{k^2}\right),\quad k\to\infty,\\[2mm]
U^+(k)= \frac{C}{R}\sqrt{\frac{c_1(v_c^2-v^2)}{M}}+o(1),\quad k\to+0,\\[2mm]
U^-(k)\sim \frac{C}{R}\sqrt{\frac{M}{c_1(v_c^2-v^2)}}\left(\frac{1}{(0+ik)^2}+\frac{S}{0+ik}\right),\quad k\to-0.
\end{gathered}
\label{Asym_Fourier}
\end{equation}

The solution of original steady-state problem \eqref{EqOfMotEta} is presented in terms of inverse Fourier transform:
\begin{equation}
u(\eta)=\frac{1}{2\pi}\int\limits_{-\infty}^{\infty} \left[U^-(k)+U^+(k)\right]e^{-ik\eta}\,dk.
\label{Solution}
\end{equation}

Constant $C$ may be defined through the asymptotic behaviour of $U^{\pm}(k)$ at infinity. With the help of Theorem A.5 in ~\cite{piccolroaz2009}, we obtain the asymptotic expression for the solution at the vicinity of the crack tip:
\begin{equation}
u(\eta)= C(1-Q\eta)+O(\eta^2),\quad \eta\to0,
\label{Asym_Solution_Zero}
\end{equation}
where the constant $Q$ is defined in \eqref{Asym_Inf_L+-}. In order to satisfy the first condition in \eqref{FractureCriterion} we must set:
\begin{equation}
C=u_c.
\end{equation}

The application of Cauchy residue theorem for the integral \eqref{Solution} with $\eta<0$ and asymptotic expression for $U^-(k)$ at zero \eqref{Asym_Fourier} reveals that:
\begin{equation}
u(\eta)\sim -\frac{C}{R}\sqrt{\frac{c_1}{M(v_c^2-v^2)}}(\eta-S),\quad \eta\to-\infty.
\label{Asym_Solution_Inf}
\end{equation}

The linear growth of the solution at $\eta\to-\infty$ arises from the assumption that the crack movement is supported by a constant energy flux from $\eta\to-\infty$. On the other hand, function $U^+(k)$ is bounded in the vicinity of $k=0$ which shows that $u(\eta)$ does not grow at $\eta\to\infty$.

From the properties of asymptotic behaviour of functions $U^{\pm}(k)$ and their regions of analyticity one may conclude that \eqref{Solution} may be reduced to:
\begin{equation}
u(\eta)=\frac{1}{2\pi}\int\limits_{-\infty}^{\infty} U^{\pm}(k)e^{-ik\eta}\,dk,\quad \pm\eta>0.
\end{equation}
Moreover, the properties of functions $L^{\pm}(k)$ and the functions in the right hand side of \eqref{WienerFactModified} give that $U^\pm(-k)=\overline{U^\pm(k)}$ for any $k\in\mathds{R}$. This leads to the following expression for the displacement:
\begin{equation}
u(\eta)=\frac{1}{\pi}\text{Re}\left(\int\limits_{0}^{\infty} U^{\pm}(k)e^{-ik\eta}\,dk\right),\quad \pm\eta>0.
\label{Solution_Computation}
\end{equation}

Finally note that the second condition in \eqref{FractureCriterion} has to be checked from the analysis of the displacement profile on the crack line ahead.
Thus in the next section we proceed to compute the full solution of the problem \eqref{Solution}. The numerical simulations of the solution profiles
below are performed utilising the formula \eqref{Solution_Computation} while the factorisation is evaluated according to the formulae \eqref{Function_l+-},\eqref{FactorizationFinal} in Appendix \ref{AppendixFactorisation} and taking into account Sokhotski-Plemelj relation.


\section{Energy release rate and displacement profile at the vicinity of a crack tip}

We set $M=1$ and consider the dimensionless values of the remaining parameters $c_1,c_2,c_3,v$ with the focus to analyse an impact of different relative values of spring stiffnesses.
For the presentation of numerical results we also set $u_c=1$.
First we present the energetic characteristics of the problem,
 that is the ratio of the global energy release rate $G$ and the local energy release rate $G_0$ (see \cite{slepyan2012}):
\begin{equation}
\frac{G_0}{G}=R^2.
\end{equation}
Note that $G_0=\frac{1}{2}c_1u_c^2$ is the strain energy accumulated in a spring with stiffness $c_1$ at the moment when the displacement reaches its critical value and
indicates the energy necessary to break a consecutive link. As it is shown in~\cite{slepyan1984}, $G$ is expressed by the sum of the local energy release rate $G_0$ and the energy carried by the elastic waves radiated from the crack tip in both directions. Hence, the less the ratio $G_0/G$ is the more energy is taken away from the crack tip by the elastic waves while the fracture occurs.

\begin{figure}[h!]
\minipage{0.33\textwidth}
a)
\center{\includegraphics[width=\linewidth]{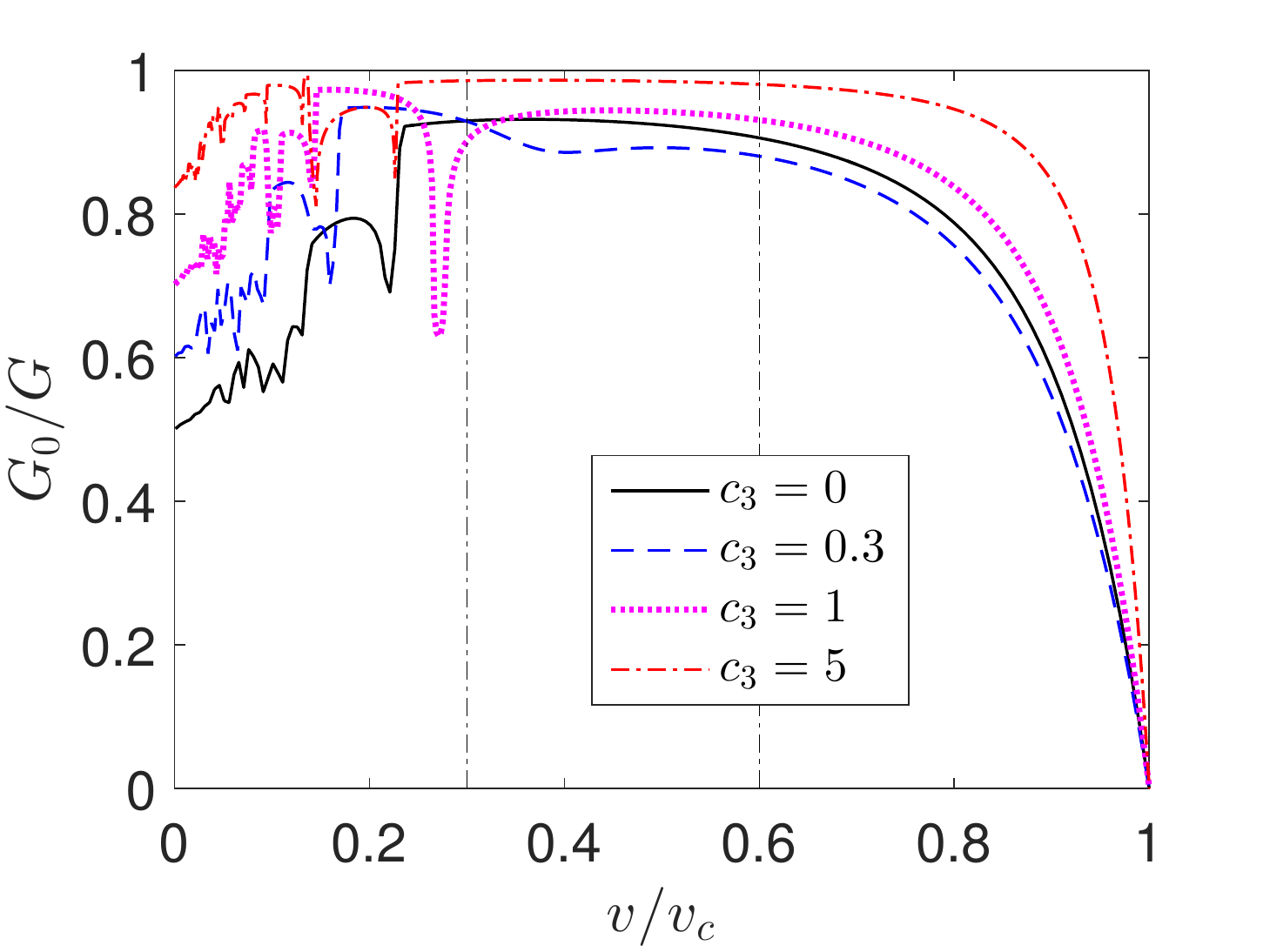}}
\endminipage
\hfill
\minipage{0.33\textwidth}
b)
\center{\includegraphics[width=\linewidth]{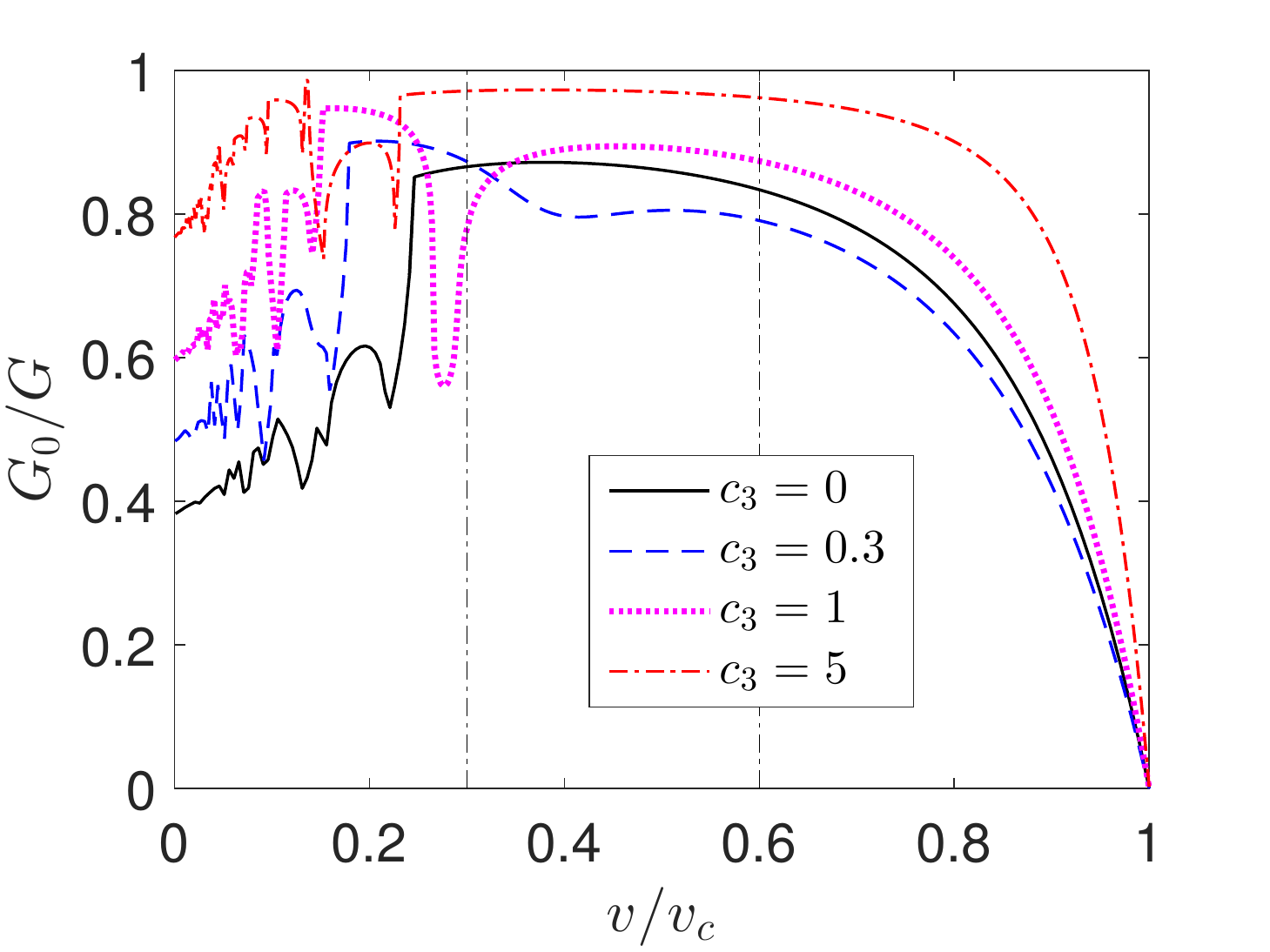}}
\endminipage
\minipage{0.33\textwidth}
c)
\center{\includegraphics[width=\linewidth]{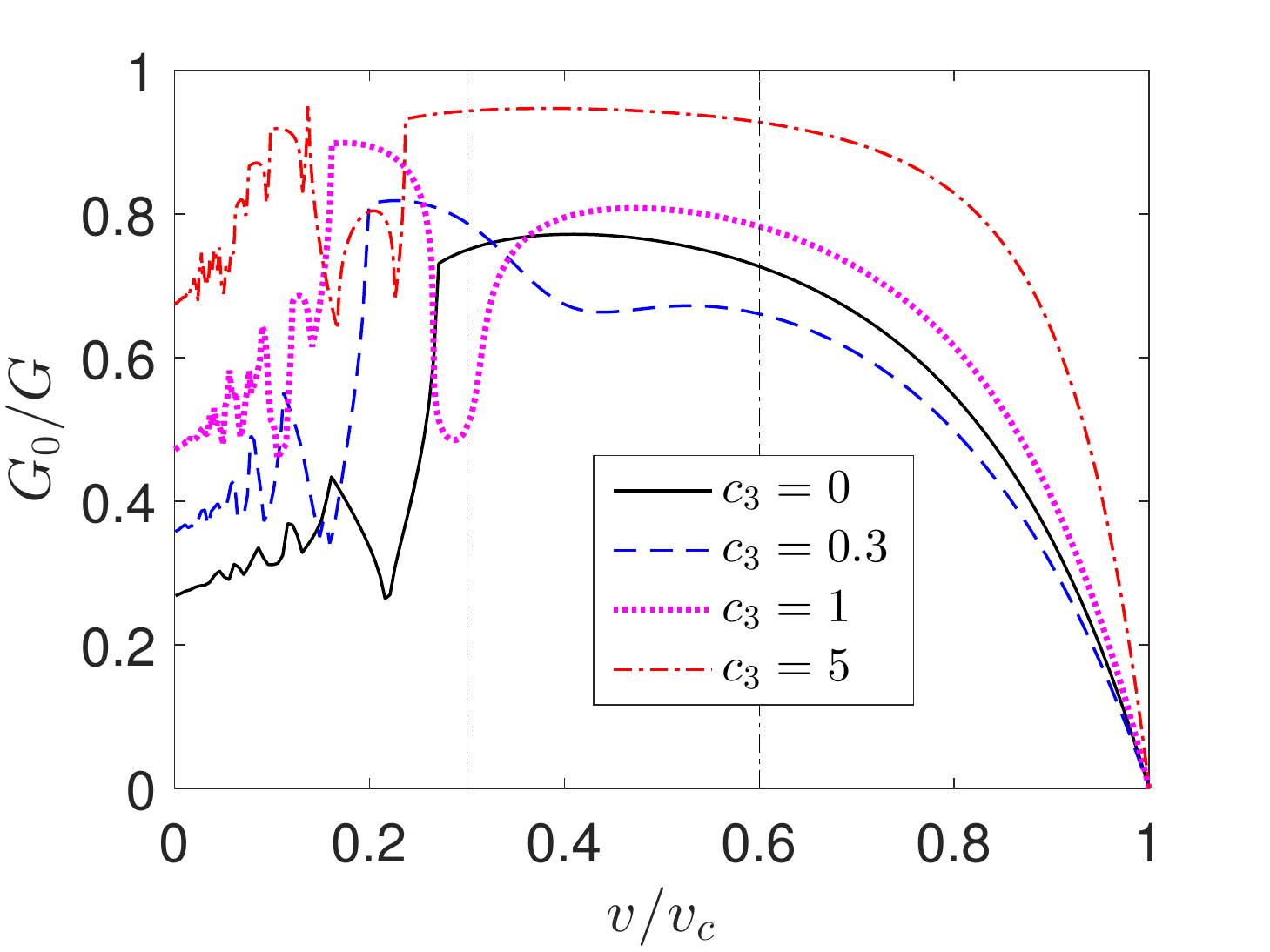}}
\endminipage
\caption[ ]{Energy release ratios $G_0/G$ with $M=1,\,c_2=1$: a) $c_1=0.5$, b) $c_1=1$, c) $c_1=2$. }
\label{EnergyDiagram}
\end{figure}

Plots of the energy release ratio $G_0/G$ for different values of parameters $c_1$ and $c_3$ and $c_2=1$ are shown on \eqref{EnergyDiagram}.
Two main distinct patterns of the $G_0/G$ -- $v$ curve are: highly oscillating behaviour of the energy ratio for relatively low values of $v$ and non-monotonous behaviour for moderate speeds and a monotonic decreasing branch for high values of the crack speed, $v$.  It is a common knowledge, starting from \cite{slepyan1984,slepyan2012,marder1995}, that these two regions reveal unstable and stable regimes of the crack propagation, correspondingly. For further details the reader is addressed to the above mentioned works.

For the same value of the crack speed, the ratio $G_0/G$ decreases with the increase of the stiffness $c_1$ responsible for the anisotropy properties of the original lattice chain with local interactions ($c_3=0$). The case of $c_1=c_2$ and $c_3=0$ corresponds to the isotropic chain with the local interactions only.

As one can expect, the energy diagrams highlight some new features attributed specifically to the structures with the introduced non-local interactions.
There is a tendency in the increase of $G_0/G$ with the increase of $c_3>0$ at high values of crack speed, approximately, $v>0.5v_c$.
Interestingly, in the absence of the non-local links ($c_3=0$), the values of $G_0/G$ of the stable branch are greater than that for $c_3=0.3$.
Moreover, with the introduced non-local interactions, there appears a drop (smooth local minimum) of the energy release rate ratio, $G_0/G$, at intermediate values of the crack speed.
This effect becomes fully suppressed when the non-local interactions are dominant (for relatively high values of $c_3$ with respect to $c_2$).
Thus, addition of the non-local interaction has a non-monotonic impact to the properties of the original chain with the local interactions only.
This potentiality gives a chance to search for some optimal properties of the new structure.

\begin{figure}[h!]
\minipage{0.33\textwidth}
a)
\center{\includegraphics[width=\linewidth]{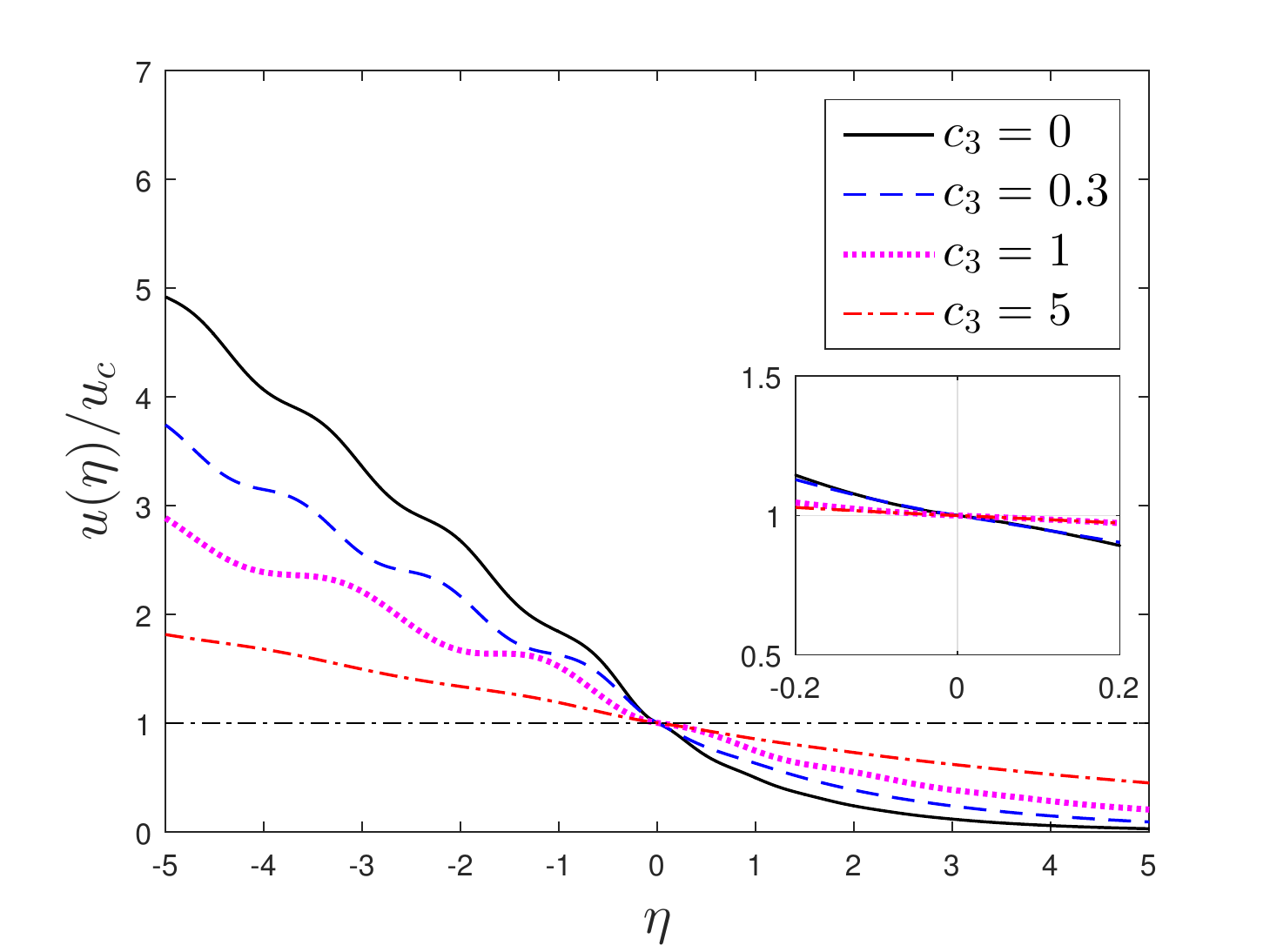}}
\endminipage
\hfill
\minipage{0.33\textwidth}
b)
\center{\includegraphics[width=\linewidth]{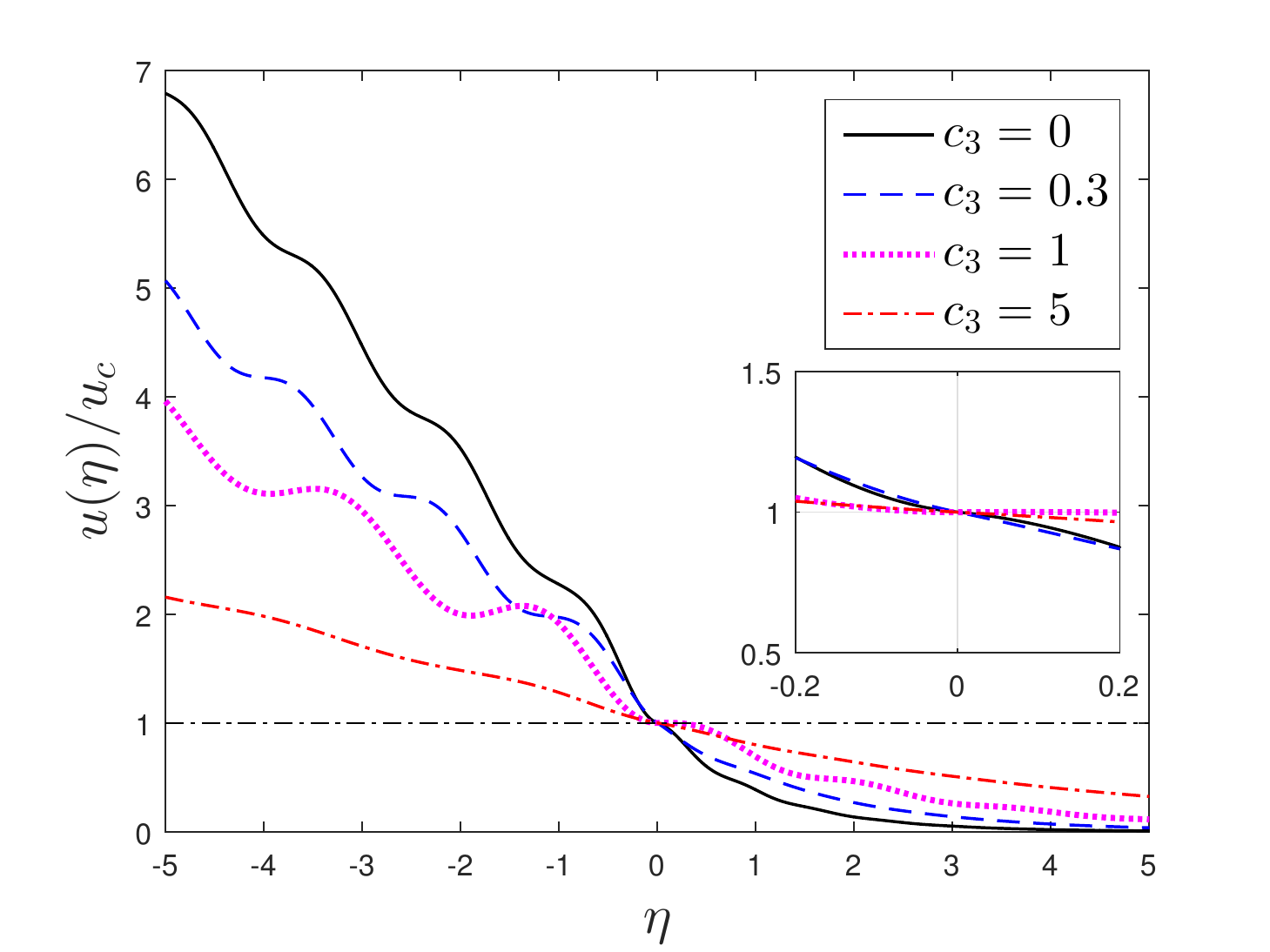}}
\endminipage
\hfill
\minipage{0.33\textwidth}
c)
\center{\includegraphics[width=\linewidth]{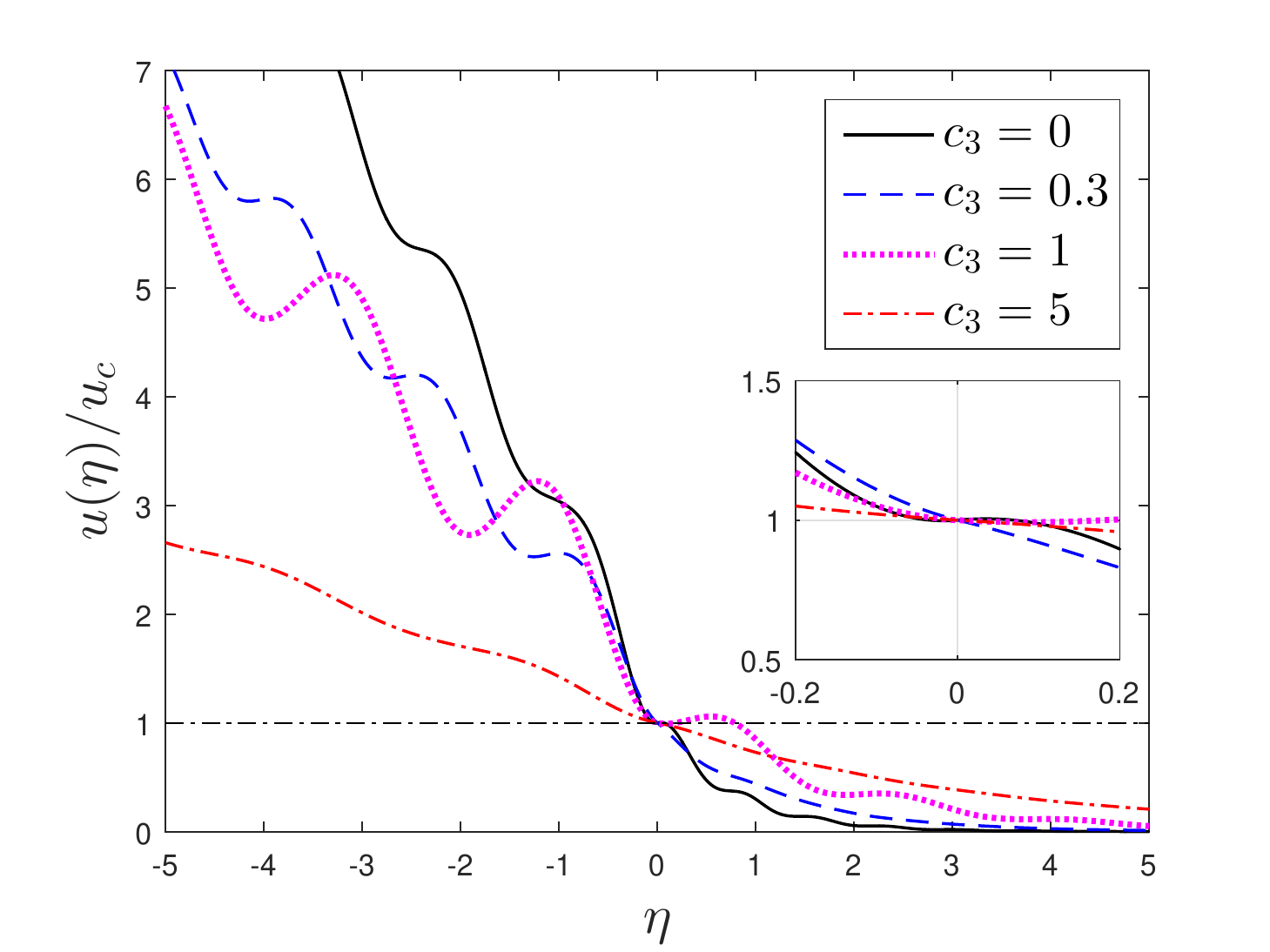}}
\endminipage
\caption[ ]{Normalised displacement $u/u_c$ at crack speed $v=0.3v_c$ with $M=1,\,c_2=1$: a) $c_1=0.5$, b) $c_1=1$, c) $c_1=2$. The plot inserts show the zoomed displacement profiles at the vicinity of a crack tip $\eta=0$.}
\label{Displacement1}
\end{figure}

Note that those local minima appearing due to the introduction of the non-local interactions should not correspond to a steady-state propagation regime as it has to violate the second condition in \eqref{FractureCriterion} for the region $\eta>0$. This is explicitly shown at the plot of the respective displacement profile on fig.\ref{Displacement1}c.

The energy diagram shows that the existence of the non-local interactions creates an additional chance to "slow down" the speed of the steadily propagating crack by applying the load in a different way as considered here. By this we mean that there is a range of values of the crack speed $v$ when it may move steadily (e.g., the small interval of values $v$ after the maximum and before smooth local minimum of $G_0/G$ curve for $c_3=0$ on fig.\ref{EnergyDiagram}). Such phenomenon for a highly anisotropic structure with the local interactions only was recently observed for a crack propagating in the structure as a result of release of internal energy ~\cite{ayzenberg2014} stored on the crack path.

\begin{figure}[h!]
\minipage{0.33\textwidth}
a)
\center{\includegraphics[width=\linewidth] {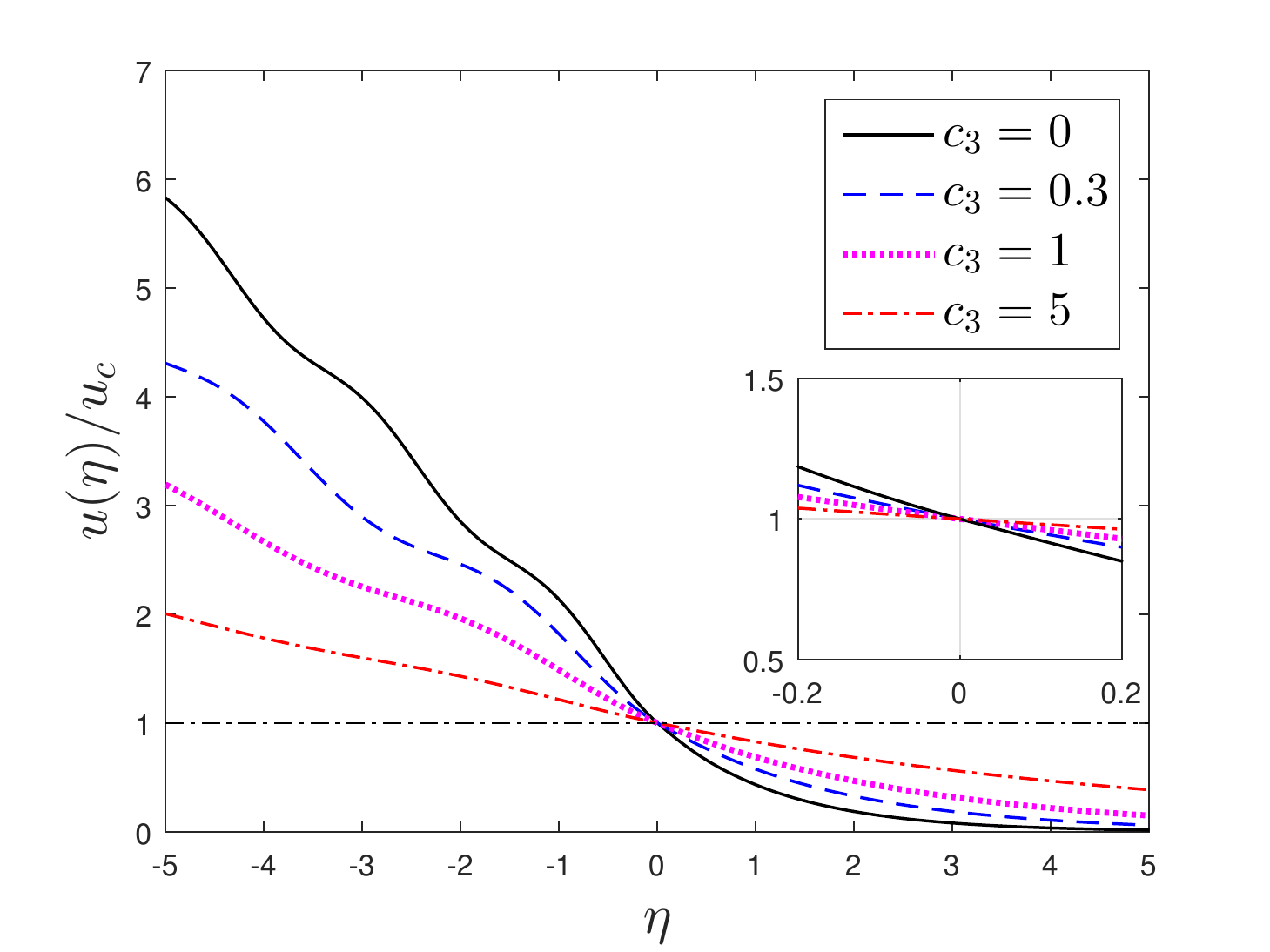}}
\endminipage
\hfill
\minipage{0.33\textwidth}
b)
\center{\includegraphics[width=\linewidth]{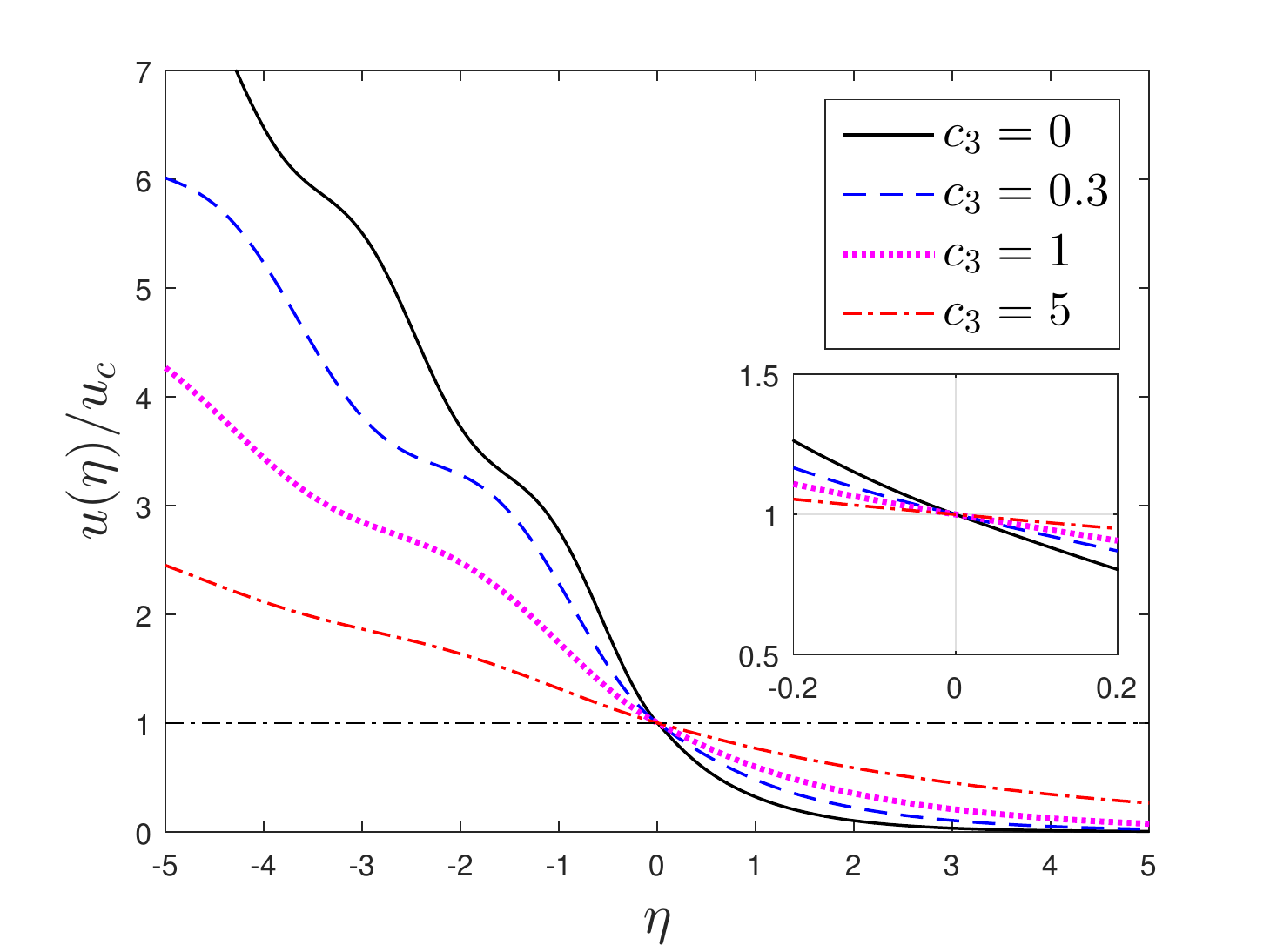}}
\endminipage
\hfill
\minipage{0.33\textwidth}
c)
\center{\includegraphics[width=\linewidth]{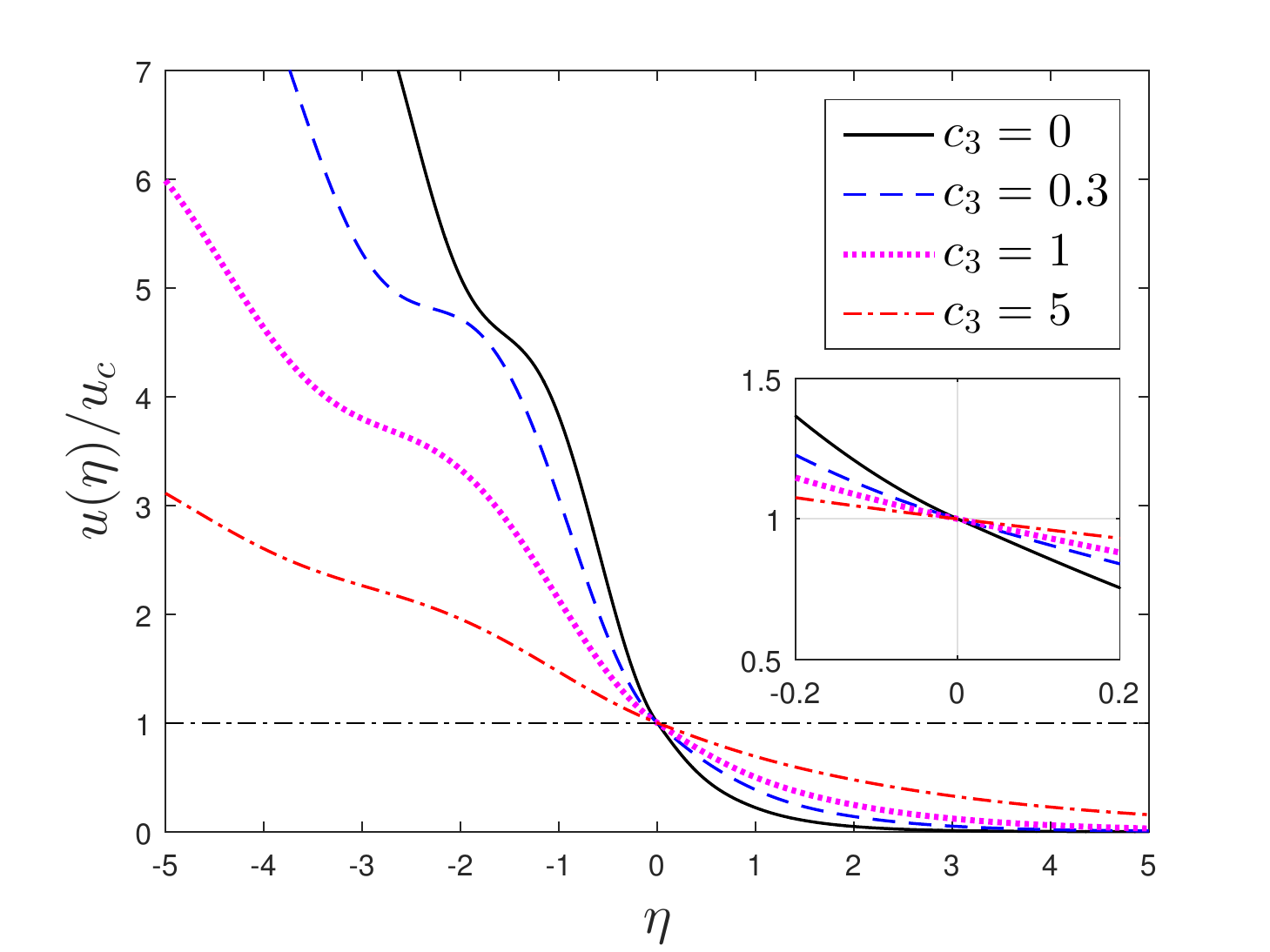}}
\endminipage
\caption[ ]{Normalised displacement $u/u_c$ at crack speed $v=0.6v_c$ with $M=1,\,c_2=1$: a) $c_1=0.5$, b) $c_1=1$, c) $c_1=2$. The plot inserts show the zoomed displacement profiles at the vicinity of a crack tip $\eta=0$.}
\label{Displacement2}
\end{figure}

Numerical results for the displacement $u(\eta)$ at crack speed $v=0.6v_c$ and $v=0.3v_c$ based on the theoretical analysis are demonstrated on fig.\ref{Displacement2}, fig.\ref{Displacement1}, respectively. We chose those values for the following reasons: when $v=0.6v_c$ the respective point on energy-speed diagram corresponds to the predicted steady-state propagation, while for speed $v=0.3v_c$ the points lie on the opposite part of the energy curve and all of them should be unstable in accordance with the prediction from~\cite{slepyan1984,slepyan2012,marder1995}. Thus we expect to observe the displacement profiles to be compatible with all the assumptions for the steady-state regime \eqref{FractureCriterion}
in the first case, and some irregularities contradicting to the steady-state regime in the opposite cases.

As it follows from the profiles presented on fig. \ref{Displacement2}, our computations fully support the expectations for the higher velocity, $v=0.6v_c$. Indeed, in this case the second condition from \eqref{FractureCriterion} is always satisfied.
In those steady-state crack propagation regimes, the displacement decreases in the broken region $\eta<0$ and increases in the intact region $\eta>0$ with the increase of $c_3$. Meanwhile, the displacement increases in the broken region $\eta<0$ and decreases in the intact region $\eta>0$ with the increase of $c_1$.
Also, no visible wave motion is observed in the broken region $\eta>0$ for the crack speed value $v=0.6v_c$,
thus in this case, most of the energy carried out by the reflected waves.

The situation changes however when we analyse the displacement profiles for the case $v=0.3v_c$.
Here one can find a collection of profiles which are also fully compatible with the steady-state propagation assumptions while should not be so in accordance with the predictions of~\cite{slepyan1984,slepyan2012,marder1995}. Indeed, for the cases $c_3=0.3$ and $c_3=5$ the point corresponding to $v=0.3v_c$ on the energy-speed diagrams lies to the left from the value of the greatest maximum of the function $G_0/G(v)$. Hence, for this case the crack propagation should be rather unstable but the displacement profiles show the opposite.

One may notice that with the increase of value $c_1$ the steady-state crack propagation becomes impossible for $c_3=1$ as the second condition from
\eqref{FractureCriterion} is violated. It is also reflected on the plots of $G_0/G$ as with the increase of $c_1$ the point $v=0.3v_c$
comes closer to the local minimum. To make it more visible a zoomed domain near the crack tip is included in the frame of the picture where the respective profiles are depicted.

The cases $c_3=1$ on fig. \ref{Displacement1}b, $c_3=1$ and $c_3=0$ on fig. \ref{Displacement1}c exhibit the profiles which are contradictory with the assumptions in \eqref{FractureCriterion}.
Indeed, in this case the crack tip location is not in the point $\eta=0$ as we assumed at the beginning but on the crack front ahead.

For the next set of plots we set the parameters of the model $c_2$ and $c_3$ in such a way that the sound speed in the broken region:
\begin{equation}
c_2+4c_3=1,\quad \left(v_c=\sqrt{\frac{c_2+4c_3}{M}}=1\right).
\label{setting}
\end{equation}
In other words, the sound speed along the broken region remains the same for all the cases considered below.
This allows us to say that if the wavelength of the applied external load is long enough then corresponding one-dimensional continuum model, described by the wave equation, has the same value of a speed of sound as a parameter. However, the response of the discrete structure on the crack propagation appears to be different which emphasizes the effect of the structured media.

\begin{figure}[h!]
\minipage{0.33\textwidth}
a)
\center{\includegraphics[width=\linewidth]{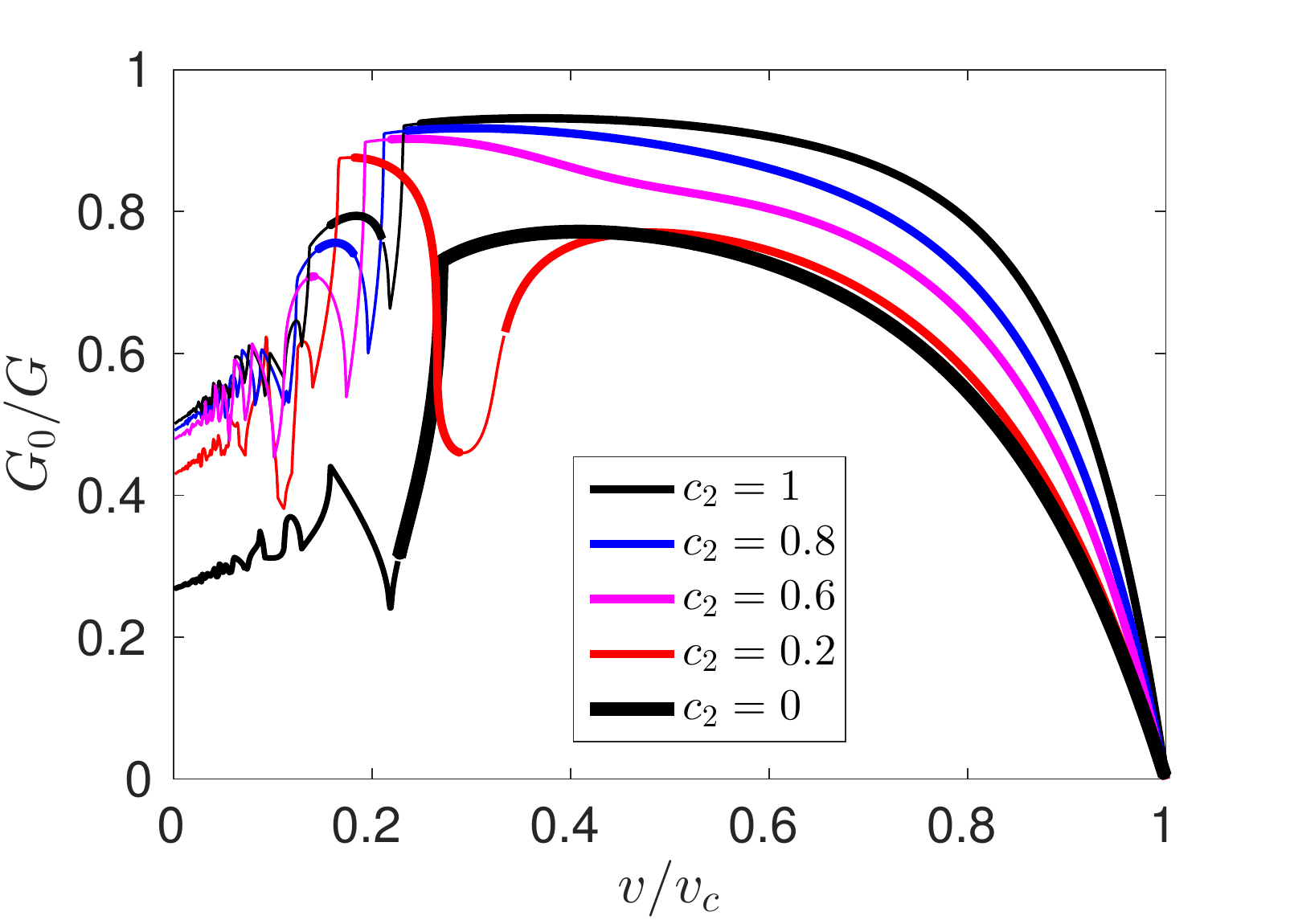}}
\endminipage
\hfill
\minipage{0.33\textwidth}
b)
\center{\includegraphics[width=\linewidth]{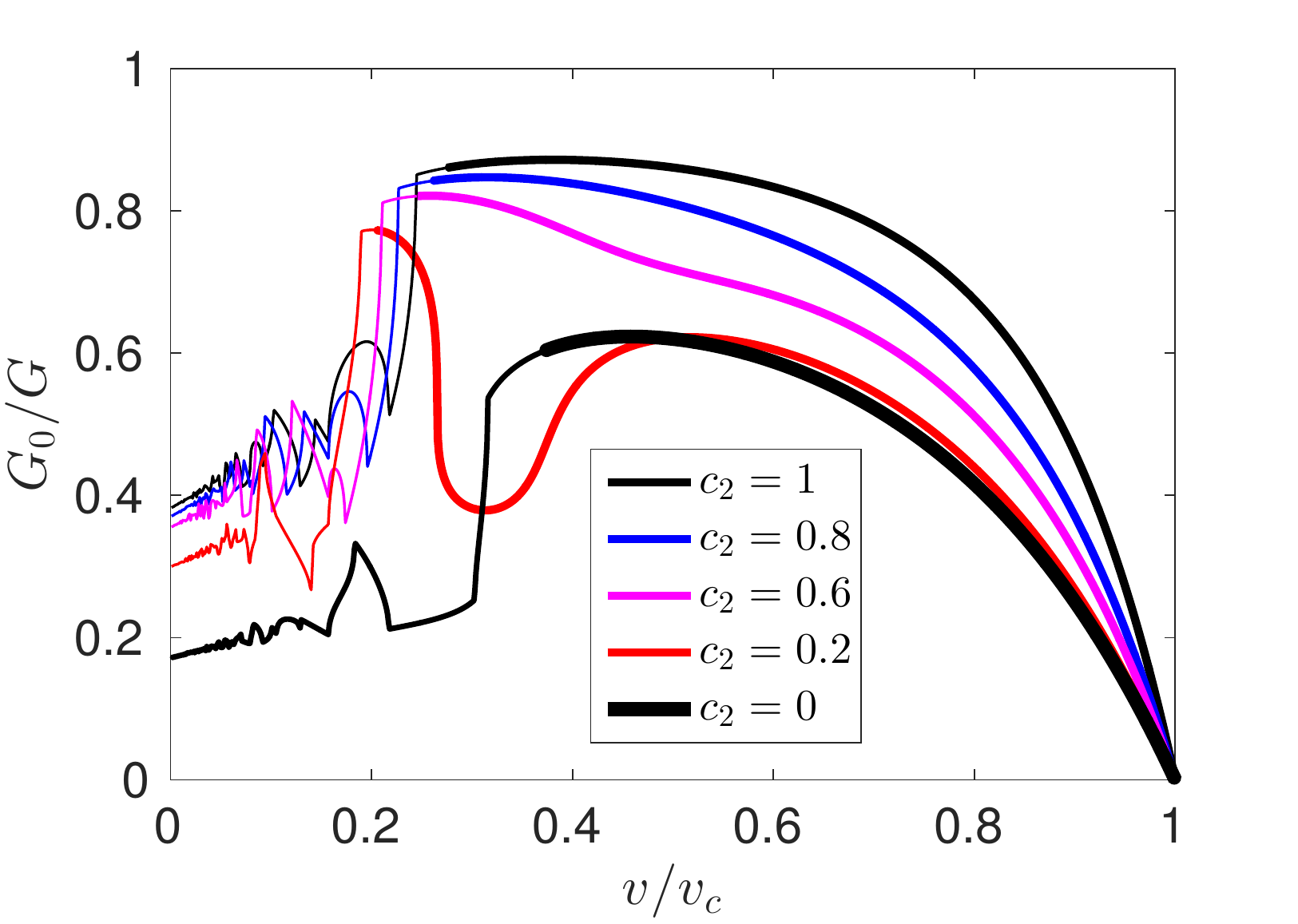}}
\endminipage
\minipage{0.33\textwidth}
c)
\center{\includegraphics[width=\linewidth]{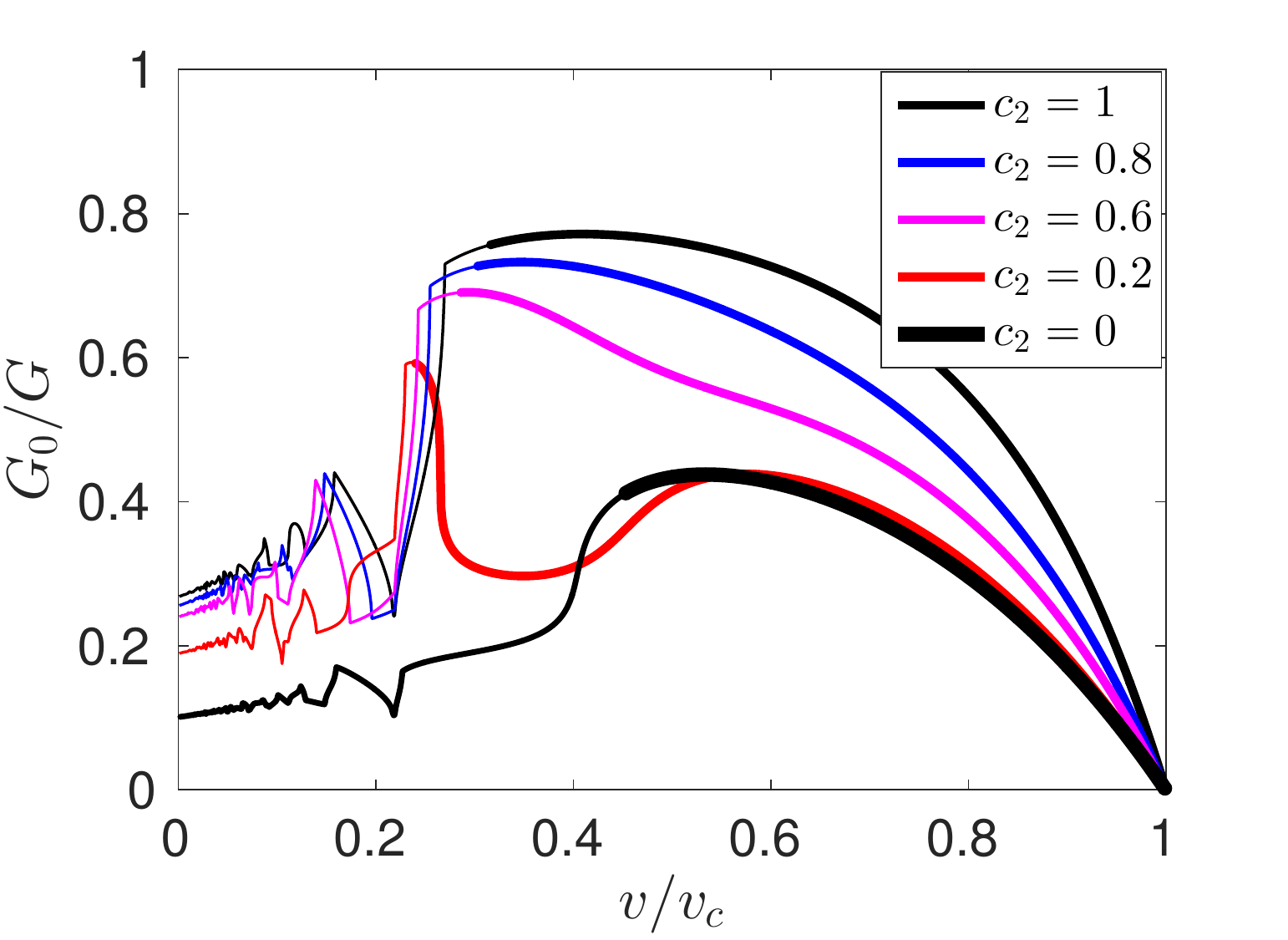}}
\endminipage
\caption[ ]{Energy release ratios $G_0/G$ for $(c_2+4c_3)/M=1$ and $M=1$: a) $c_1=0.5$, b) $c_1=1$, c) $c_1=2$. The fat lines correspond to the values of crack speed for which the condition \eqref{FractureCriterion} is fulfilled.}
\label{EnergyDiagramSameSoundSpeed}
\end{figure}

The corresponding energy - speed diagrams are presented on fig.\ref{EnergyDiagramSameSoundSpeed} for the different values of $c_1$. The range of different values of $c_2$ in accordance with \eqref{setting} shows the effect of introduced non-local interactions from the absence of them ($c_2=1$) to their absolute dominance ($c_2=0$). The latter case corresponds to the configuration of the structure with local interactions when the equilibrium lengths of the springs in horizontal direction is equal to $2a$ whereas in vertical equilibrium lengths equal to $a$. We studied the displacement field for each case and plotted the curves $G_0/G$ with fat lines for the ranges of crack speed for which the condition \eqref{FractureCriterion} is met.

As one can observe from the energy-speed diagrams in the chosen setting \eqref{setting} the increase of the parameter $c_2$ (the respective decrease of $c_3$) impacts the behaviour of the curve monotonically in the high speed region in contrast with the previous normalisation (when we fixed the parameters of the locally interacting springs $c_1$ and $c_2$). In accordance with the predictions, the limiting cases $c_2=1$ and $c_2=0$ do not exhibit any local smooth minima on their related energy-speed diagrams. At the same time, these minima appear with the interplay between the local and non-local interactions.

Finally, we would like to analyse the aforementioned phenomena of existence steady-state crack propagation regimes on the parts of the energy-speed diagrams where they should not exists in accordance with the predictions from~\cite{slepyan1984,slepyan2012,marder1995}.

For this, we choose the case $M=1,c_1=0.5,c_2=c_3=0.2$ and consider two levels of the energy ratios: $G_0/G=0.53$ and $G_0/G=0.7$. The reason to select those value was to consider different numbers of possible states with the points lying at different parts of the diagrams.

\begin{figure}[h!]
\minipage{0.33\textwidth}
a)
\center{\includegraphics[width=\linewidth]{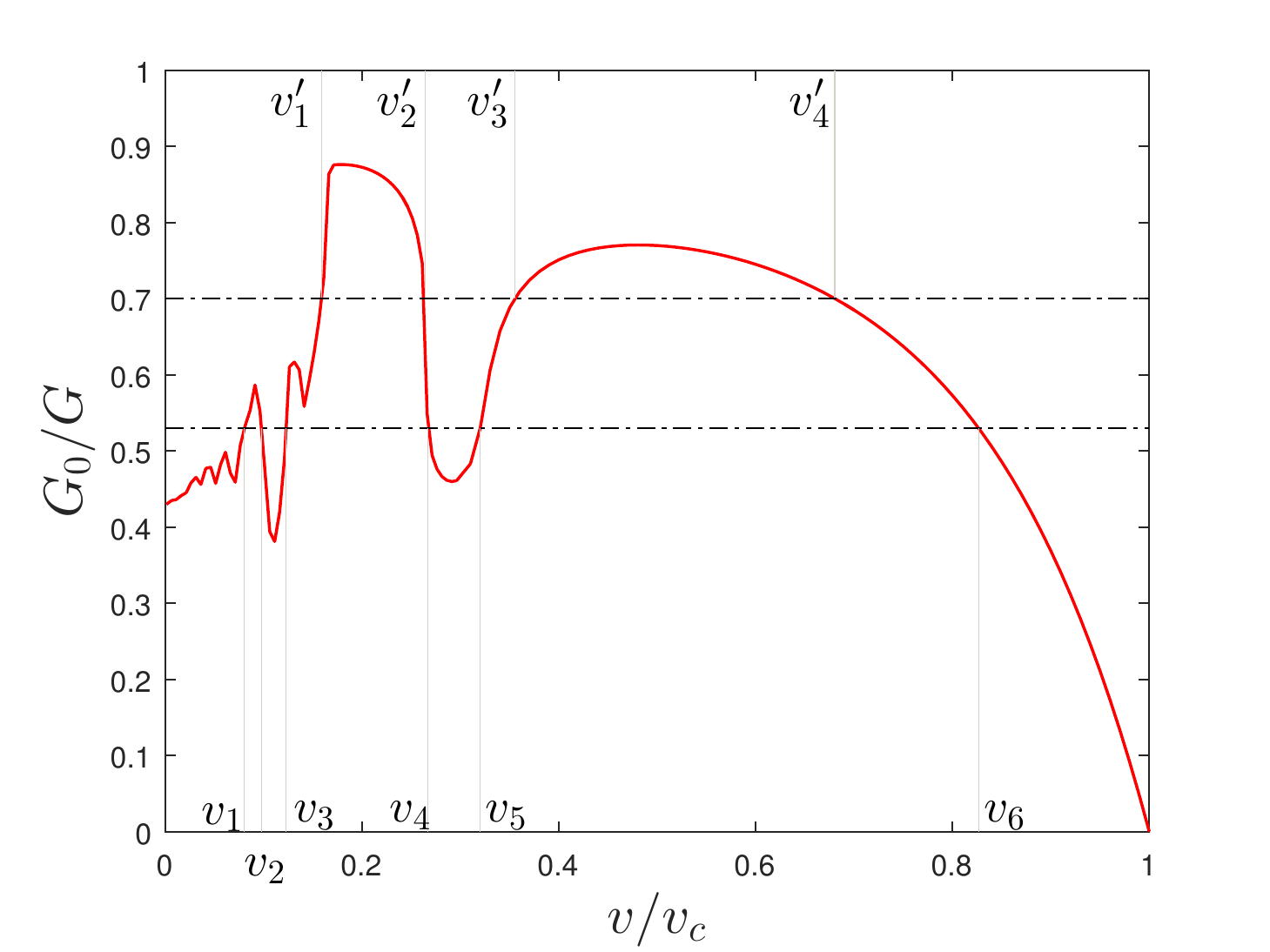}}
\endminipage
\hfill
\minipage{0.33\textwidth}
b)
\center{\includegraphics[width=\linewidth]{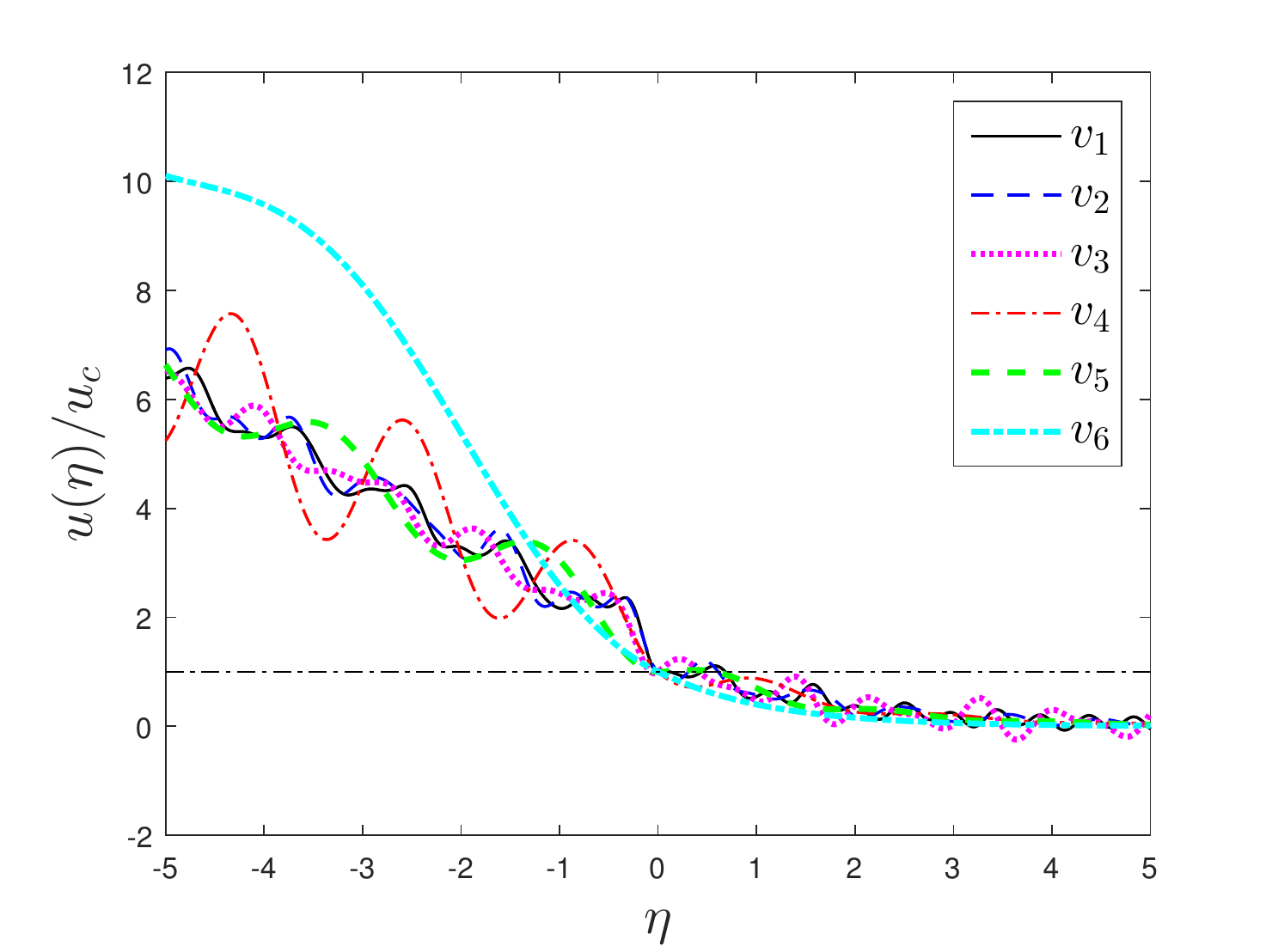}}
\endminipage
\minipage{0.33\textwidth}
c)
\center{\includegraphics[width=\linewidth]{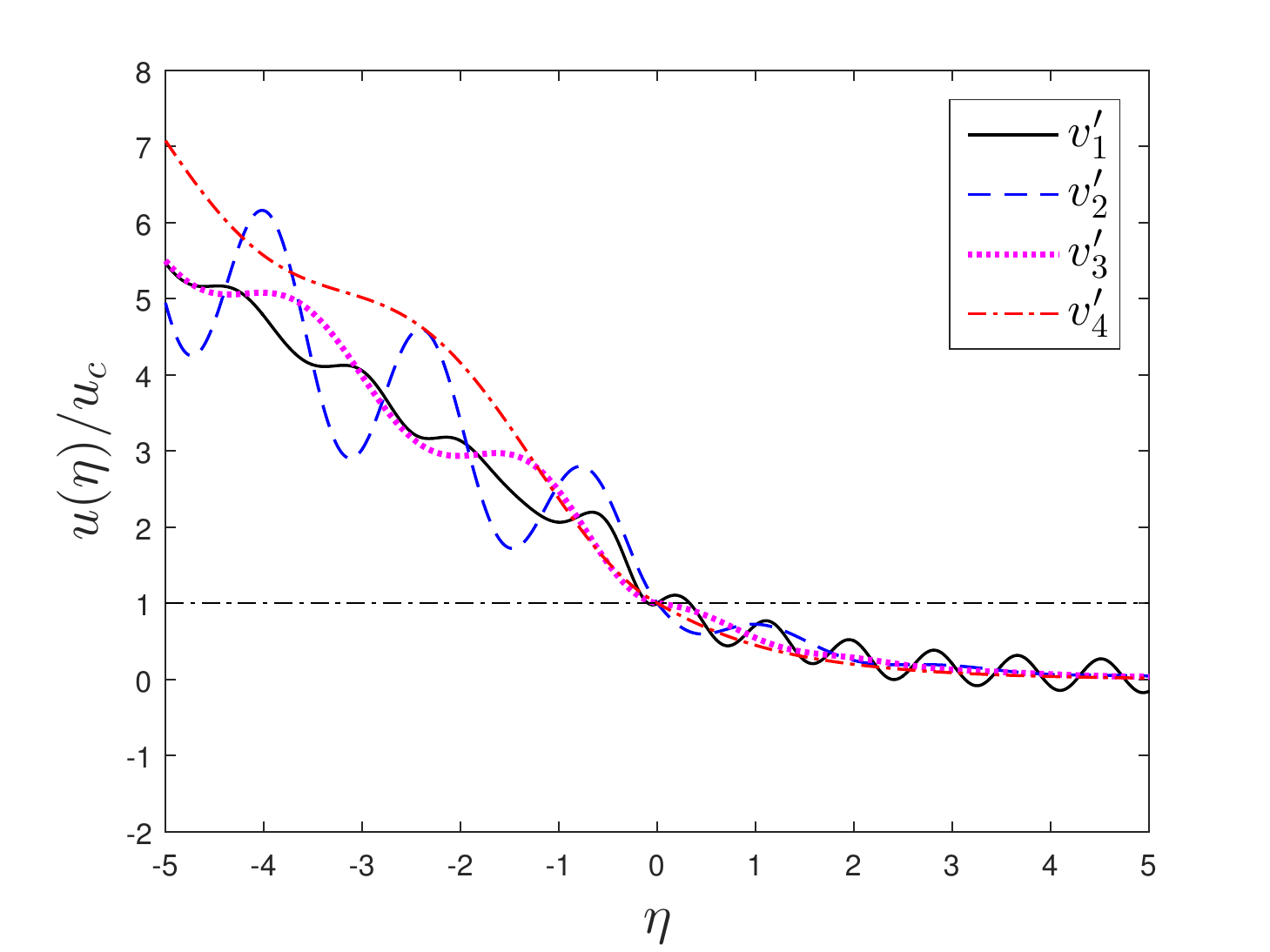}}
\endminipage
\caption[ ]{The case for when the parameters of the system $M=1$ and $c_1=1,c_2=c_3=0.2$. Values $v_1,...,v_6$ on each plot correspond to the level $G_0/G=0.53$ while $v_1',...,v_4'$ on each plot correspond to the level $G_0/G=0.7$: a) Energy release ratio $G_0/G$, b) Displacement profile $u(\eta)/u_c$ at energy release ratio level $G_0/G=0.53$, c) Displacement profile $u(\eta)/u_c$ at energy release ratio level $G_0/G=0.7$.}
\label{SameR2}
\end{figure}

For the value of energy release ratio $G_0/G=0.53$ we found the corresponding values of the crack speed $v_1,...,v_6$ which are shown on fig.\ref{SameR2}a.
We determined the respective values of the speed from the energy release ratio curve $G_0/G$ as $v_1=0.08015v_c$, $v_2=0.097285v_c$, $v_3=0.12272v_c$, $v_4=0.26687v_c$, $v_5=0.31994v_c$, $v_6=0.8265v_c$. Here the speed of sound, as previously, $v_c=1$.
We also analyse the case $G_0/G=0.7$ when four possible crack speed are: $v_1'=0.15875451v_c$, $v_2'=0.2640065v_c$, $v_3'=0.3550686v_c$, $v_4'=0.680595v_c$.

\begin{figure}[h!]
\minipage{0.5\textwidth}
a)
\center{\includegraphics[width=\linewidth]{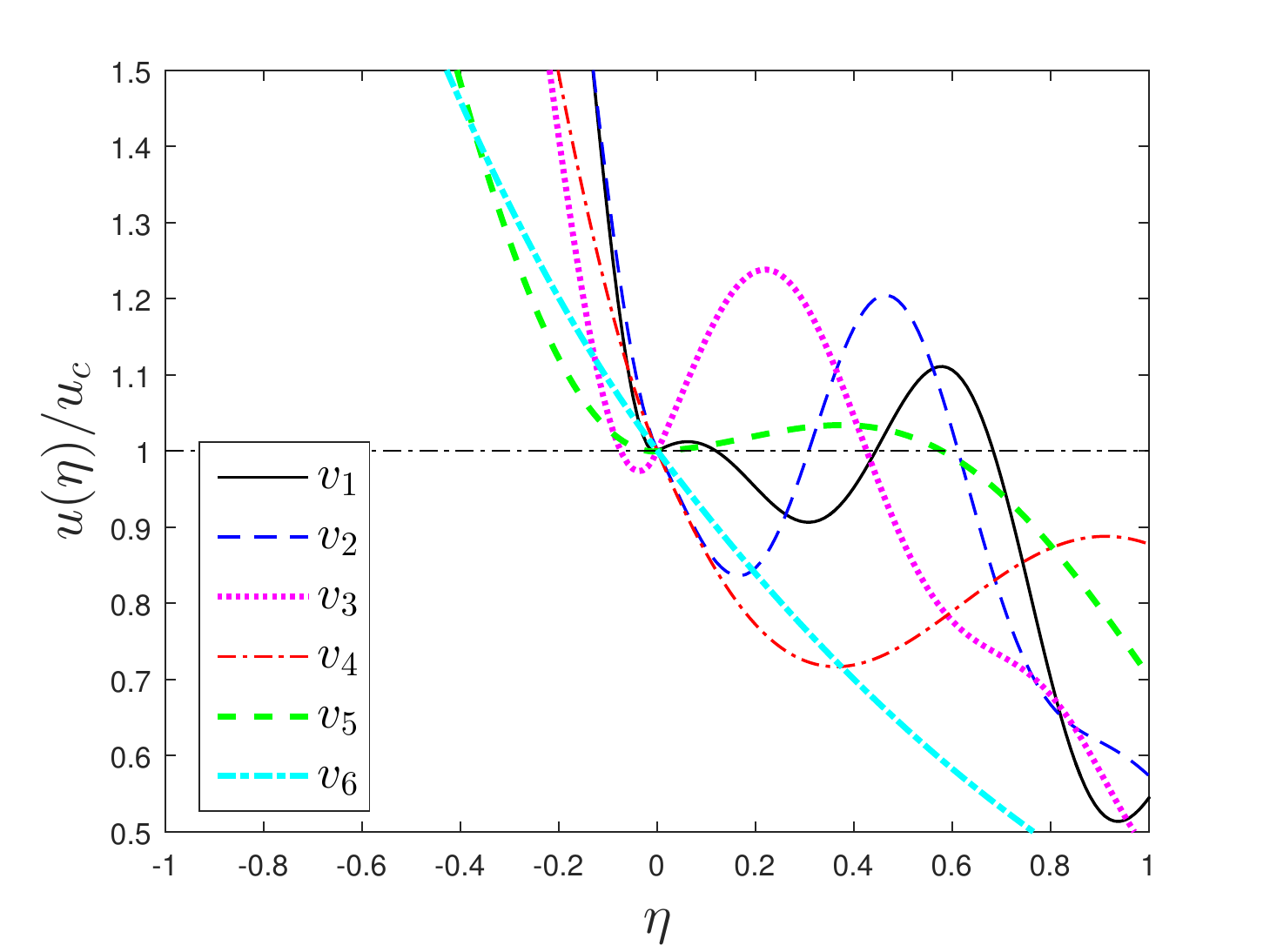}}
\endminipage
\minipage{0.5\textwidth}
b)
\center{\includegraphics[width=\linewidth]{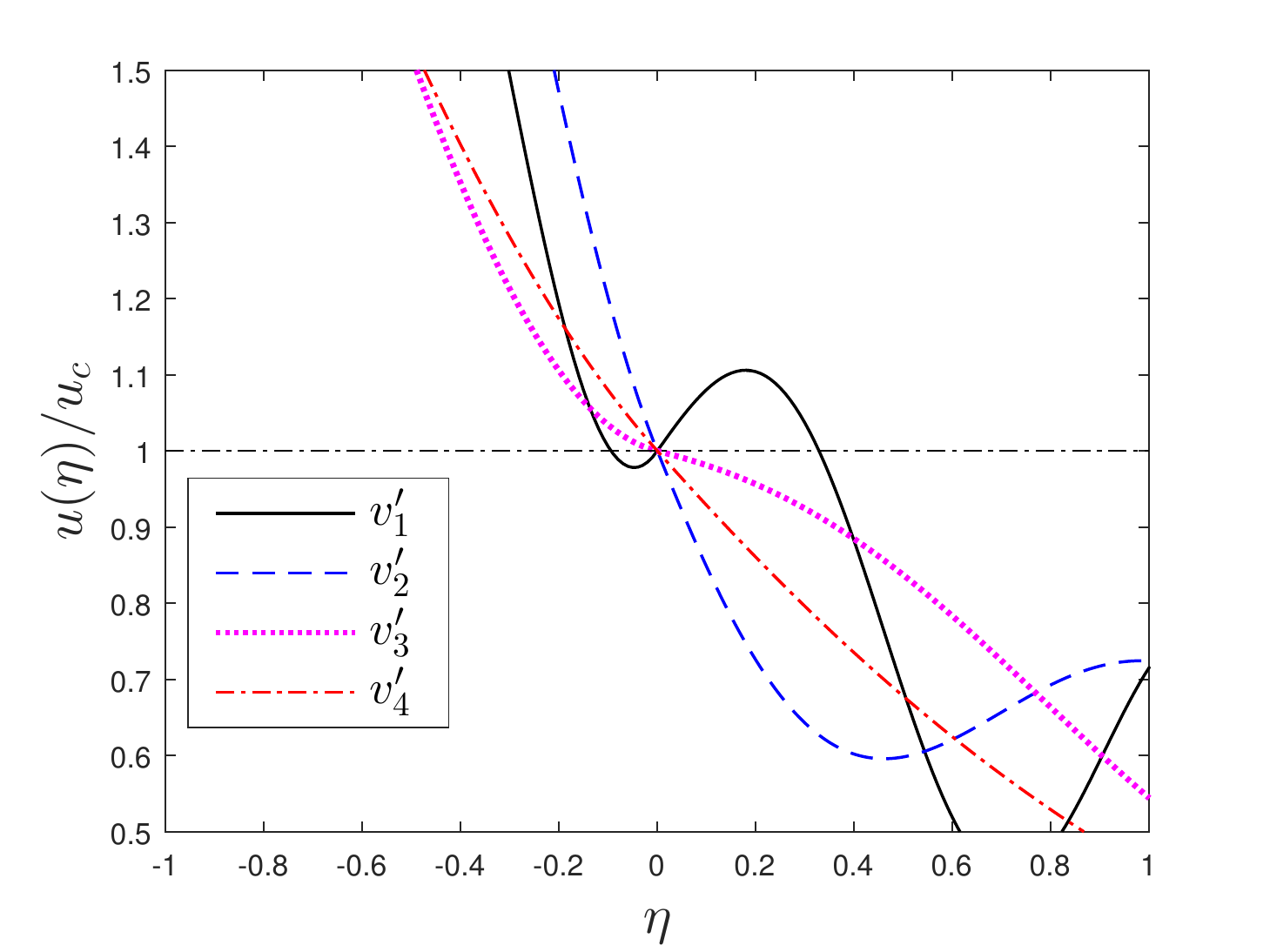}}
\endminipage
\caption[ ]{The displacement profiles at the vicinity of a crack tip. The case for when the parameters of the system $M=1$ and $c_1=1,c_2=c_3=0.2$. Values $v_1,...,v_6$ on each plot correspond to the level $G_0/G=0.53$ while $v_1',...,v_4'$ on each plot correspond to the level $G_0/G=0.7$: a) Displacement profile $u(\eta)/u_c$ at energy release ratio level $G_0/G=0.53$, b) Displacement profile $u(\eta)/u_c$ at energy release ratio level $G_0/G=0.7$.}
\label{SameR2Zoomed}
\end{figure}

According to the expectations from~\cite{slepyan1984,slepyan2012,marder1995}, one case for each energy level, $G_0/G=0.53$ and $G_0/G=0.7$, has to indicate the steady-state regime,$v=v_6$ and $v=v_4'$, respectively. From the previous discussion one may also expect that the case $v=v_4$ for $G_0/G=0.53$ and $v=v_2'$ for $G_0/G=0.7$ may correspond to the steady-state regime. The remaining cases should demonstrate the instability in crack movement. In order to draw some conclusions we analyse the displacement profiles for all these cases and present them on
fig.\ref{SameR2}b and fig.\ref{SameR2}c while the zoomed plots at the vicinity of the crack tip are shown at fig.\ref{SameR2Zoomed}a and fig.\ref{SameR2Zoomed}b, respectively.

It turns out that for the cases $v=v_4$ and $v=v_6$ for $G_0/G=0.53$ are coherent with the assumptions in \eqref{FractureCriterion} while the case $v=v_3'$ for $G_0/G=0.7$ does not obey it.

\section{Conclusions}

In the present paper we considered steady-state crack propagation in the chain structure with the non-local (next to nearest neighbour) interactions which are modelled by
linear springs of different stiffness. Accurate analytical analysis of the problem was done by utilizing Wiener-Hopf technique.
Evaluation of the displacement profiles of the structure in the steady-state regime required numerical evaluation of the Hilbert and Fourier transforms,
while the asymptotic analysis of their images allowed us to obtain the expression for the diagrams linking the energy release rate with the crack speed.

The results revealed a distinct behaviour of the energy curves with change of the material parameters.
As expected there is always a stable part of the diagrams in the high speed region and an unstable part of the diagram for a slow speed.
The major difference, in comparison with discrete structures constructed by the local interactions only, is the rather significant influence of the
non-local interactions leads to moderate crack speeds where an additional smooth local minimum and maximum can appear on the energy diagram and its size is directly predefined by the
interplay between the local and non-local interactions. If the material parameters defining the local interactions are fixed, introduction of additional non-local interactions in the structure may impact which is not monotonous. Interestingly, a complete replacement of the local links by the next to the neighbour ones completely suppress the phenomenon.
This increases a possibility to develop lattice structures with presence of both local and non-local interactions which are optimized with respect
to chosen properties of the structures.

To decide whether the steady-state movement of the crack for a specific level of external energy and given crack speed is possible or impossible we performed
very accurate computations of the displacement profiles with a special focus on the profiles on crack front ahead of the tip. In some cases, the
profile in a very small vicinity of the crack tip can only prove nonexistence of the steady-state regime.
The following new phenomena have been observed in the analysis that should be highlighted in this context:
\begin{itemize}
\item there may exist steady-state regimes corresponding to the points situated on the increasing branches of the energy -- speed diagrams on the left hand side from the last
local maximum of the diagram,
\item there may exist steady-state regimes corresponding to the points situated on the decreasing regions of the energy -- speed diagrams for low values of crack speeds.
\end{itemize}
However, these conclusions are not decisive and to prove the existense/nonexistence of the regimes further extensive analysis is required.
Numerical simulations directly performed for the original discrete system should be a part of such analysis.

\section*{Acknowledgments}

NG and GM were supported by the FP7 PEOPLE Marie Curie ITN project CERMAT2 under the number \textit{FP7-MC-ITN-2013-606878-CERMAT2}.
\bibliographystyle{abbrv}
\bibliography{Bibliography}



\appendix
\section{Relation between dispersion relations and function $L(k)$.}
\label{AppendixDispersion}

First we analyse how waves can propagate along the two different parts of the chain: the broken and the intact ones.
As usual, we consider a possible solution in the form of a harmonic wave $u_n(t)=e^{i(kn-\omega{t})}$.
Substituting this representation into the respective parts of the equations \eqref{EquationOfMotion}
we get dispersion relations for the intact region of chain ($\omega_1(k)$) and the broken one ($\omega_2(k)$):
\begin{equation}
\begin{gathered}
\omega_1^2(k)=\frac{4}{M}\left[c_3\sin^2{k}+c_2\sin^2\left(\frac{k}{2}\right)+\frac{c_1}{4}\right], \quad n\geq{n^*}\\
\omega_2^2(k)=\frac{4}{M}\left[c_3\sin^2{k}+c_2\sin^2\left(\frac{k}{2}\right)\right], \quad  n<n^*
\end{gathered}
\label{Omega_1_2}
\end{equation}

\begin{figure}[h!]
\begin{minipage}[h]{0.5\linewidth}
a)
\center{\includegraphics[width=1\linewidth]{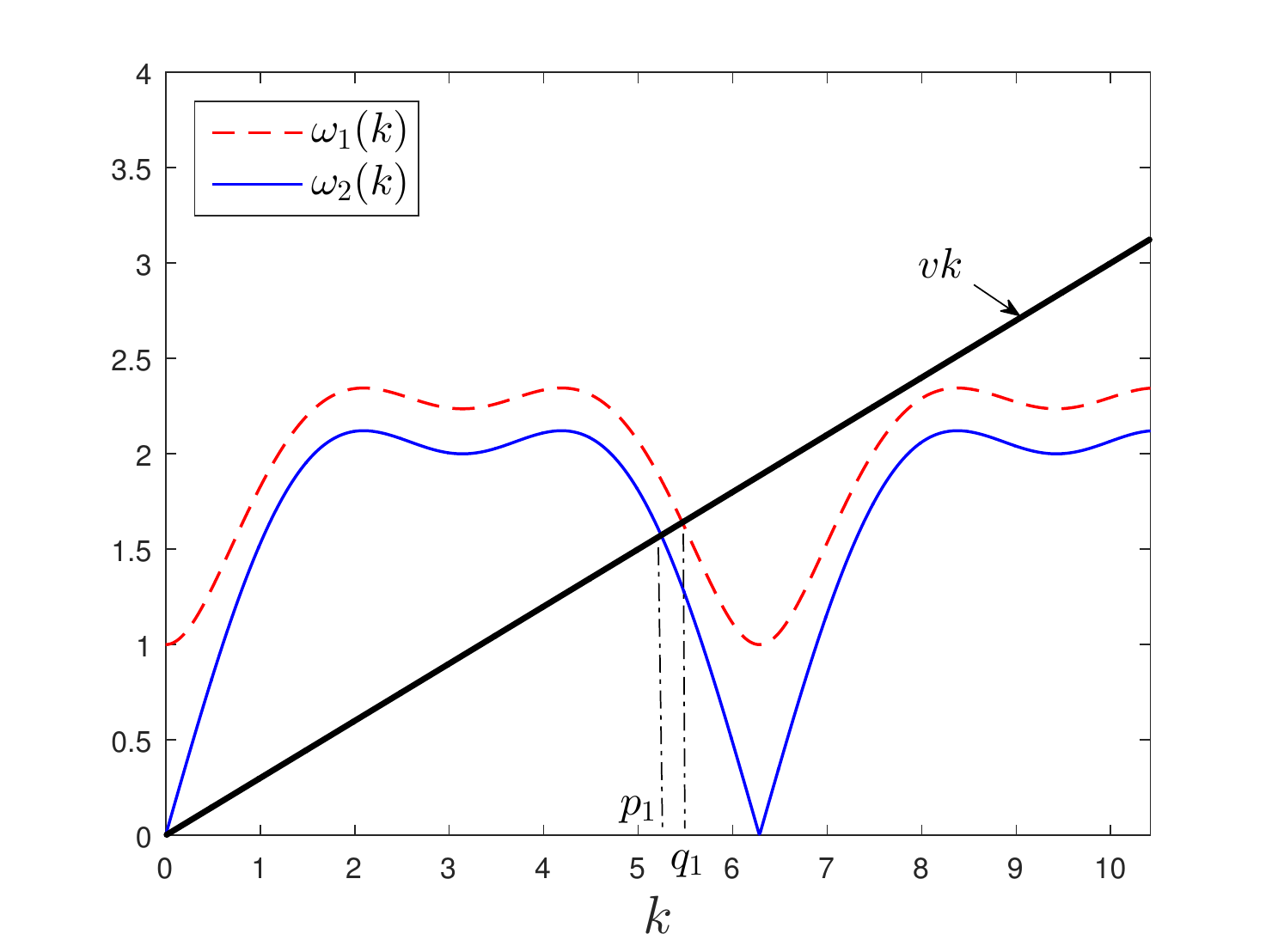}}
\end{minipage}
\hfill
\begin{minipage}[h]{0.5\linewidth}
b)
\center{\includegraphics[width=1\linewidth]{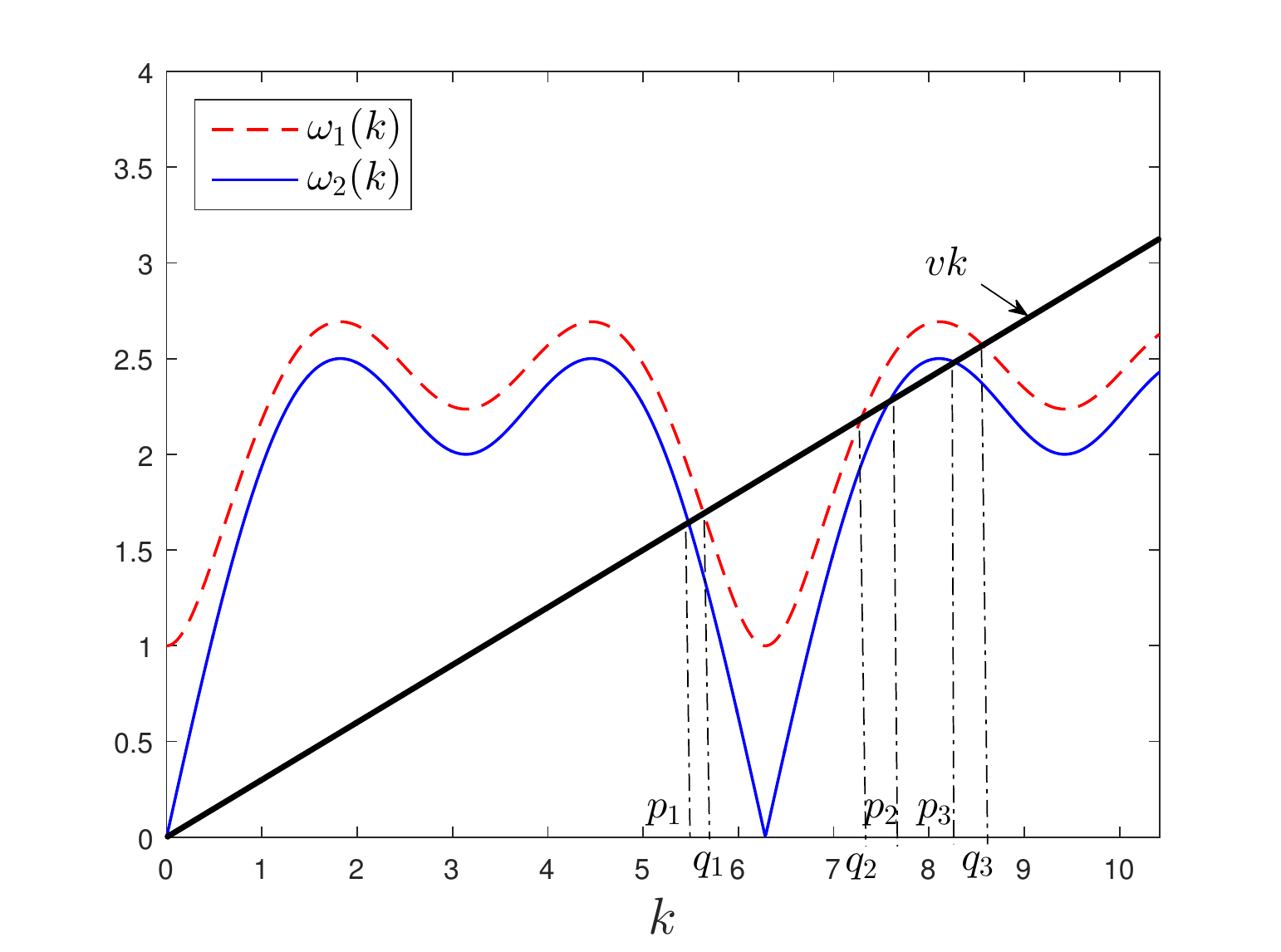}}
\end{minipage}
\caption[ ]{ Dispersion diagram of a chain with non-local interactions for $M=1,c_1=c_2=1,v=0.3$. The relation $\omega_1(k)$ corresponds to an intact region whereas $\omega_2(k)$ to broken one. The different figures stand for the following values of $c_3$: a) $c_3=0.5$, b) $c_3=1$. }
\label{DispersionDiagram}
\end{figure}

With \eqref{Omega_1_2} one can notice that the function $L(k)$ defined in \eqref{FunL} and which has to be factorized is:
\begin{equation}
L(k)=\frac{(0+ikv)^2+\omega_1^2(k)}{(0+ikv)^2+\omega_2^2(k)},
\label{FunLDisp}
\end{equation}

Thus, the factorisation of function $L(k)$ is directly related with its zeros and poles along the real line. These real zeros and poles are the roots of the equations:
\begin{equation}
\omega_{j}^2(k)-(vk)^2=0,\quad j=1,2
\label{eqn1}
\end{equation}

From the definition of $\omega_{1,2}(k)$ \eqref{Omega_1_2} we conclude that these functions are even for $k\in\mathds{R}$. As a result, if there is a real root of the equation (\eqref{eqn1}) $k=k_*^{j}$ ($j=1,2$) then $k=-k_*^{j}$ ($j=1,2$) is also a solution of the equation.

For the fixed value of the crack speed $v$ let us introduce following notations:
\begin{equation}
\begin{gathered}
q_j>0:\quad \omega_1(q_j)-vq_j=0,\quad j=1, \ldots,Z,\\
p_j>0:\quad \omega_2(p_j)-vp_j=0,\quad j=1,\ldots,P,
\end{gathered}
\label{ZerosPoles}
\end{equation}
with $P$ and $Z$ being integers. One can see the plots of normalized frequencies $\omega_{1,2}(k)$ and above mentioned intersection points on fig.\ref{DispersionDiagram} for different values of parameters $c_3$. There is also a root at point $k=0$ of function $\omega_2(k)$:
\begin{equation}
\omega_2(0)=0,
\end{equation}

\section{Factorization of function $L(k)$}
\label{AppendixFactorisation}

Apart from the fact that factorization of the function $L(k)$ from \eqref{FunL} satisfying an additional normalization condition at infinity is unique,
it can be performed by various techniques not only in the way suggested by \eqref{Factorization}.
One should also remember that the inverse Fourier transform of the solution
should be performed in order to evaluate the displacement profiles. Moreover,
information on irregular points of the function along the real axis help enormously during the computation of the inverse operator.
For this reason, we separate the real zeros and poles of the function $L(k)$ and also conserve the way how the integration path behaves to avoid the points
(from below or above). Let us express the function $L(k)$ as:
\begin{equation}
L(k)=L_0(k)l^+(k)l^-(k),
\end{equation}
where function $L_0(k)$ possesses no zeros or poles along the real axis and $|L_0(k)|\to1$ as $k\to\infty$. To achieve this,
we define the functions $l^{\pm}(k)$ in the following manner:
\begin{equation}
\begin{gathered}
l^-(k)=\frac{(p_0+ik)^{d+1}}{0+ik}\frac{\displaystyle\prod_{j=1}^{Z'}(0+i(k-q_{2j-1}))(0+i(k+q_{2j-1}))}{\displaystyle\prod_{j=1}^{P'}(0+i(k-p_{2j-1}))(0+i(k+p_{2j-1}))},\\
l^+(k)=\frac{(p_0-ik)^{d+1}}{0-ik}\frac{\displaystyle\prod_{j=1}^{Z'}(0-i(k-q_{2j}))(0-i(k+q_{2j}))}{\displaystyle\prod_{j=1}^{P'}(0-i(k-p_{2j}))(0-i(k+p_{2j}))},\\
d=P-Z,\,Z'=\Big{\lceil}\frac{Z+1}{2}\Big{\rceil},\,P'=\Big{\lceil}\frac{P+1}{2}\Big{\rceil}.
\end{gathered}
\label{Function_l+-}
\end{equation}
Here, $P$ is the number of real, positive poles of function $L(k)$ in limit $s\to0+$ (positive roots of equation $\omega_1(k)-vk=0$),
whereas $Z$ is the number of real, positive zeros of function $L(k)$ (positive roots of equation $\omega_2(k)-vk=0$).
Symbol $\lceil x \rceil$ stands for the integer part of a number $x$.

In order to decide whether there is one or more multipliers of the numerator or denominator of $l^{\pm}(k)$ one
should consider the equations $\omega_{j}^2(k)-(s+ikv)^2=0,$ $j=1,2$ at the vicinity of its root $k=k^*$ in the limit case $s\to+0$
and to take into account that $v<v_c$. The reader may also refer to~\cite{slepyan2012} for the details.
Real constant $p_0>0$ is chosen in such a way that the condition $L_0(0)=1$ is satisfied when $s\to+0$ , or explicitly:
\begin{equation}
p_0=\left(\frac{c_1}{-Mv^2+4c_3+c_2}\left(\prod_{j=1}^{P}p_j^2/\prod_{j=1}^{Z}q_j^2\right)\right)^\frac{1}{2d+2}.
\end{equation}

Functions $l^{\pm}(k),L_0(k)$ inherit the properties of the function $L(k)$ (see \eqref{PropertiesL}) and
thus $L_0(k)$ can be factorized in the way shown previously in \eqref{Factorization}.
The final factorization of the function $L(k)$ can be expressed via $l_{\pm}(k)$ and $L_{0\pm}(k)$:
\begin{equation}
\begin{gathered}
L(k)=L^+(k)L^-(k),\\
L^{\pm}(k)=l^{\pm}(k)L_0^\pm(k).
\end{gathered}
\label{FactorizationFinal}
\end{equation}

\end{document}